\documentclass[twocolumn, tighten]{aastex63}
\usepackage{mathtools}
\usepackage{enumitem}
\usepackage{appendix}

\graphicspath{{./}{figures/}}
\begin{document}

\title{A Morphological Study on Galaxies Hosting Optical Variability-Selected AGNs in the COSMOS Field}

\author{Yuxing ZHONG}
\affiliation{Department of Pure and Applied Physics, Graduate School of Advanced Science and Engineering, Faculty of Science and Engineering, Waseda University, 3-4-1 Okubo, Shinjuku, Tokyo 169-8555, Japan}

\author{Akio K. Inoue}
\affiliation{Department of Pure and Applied Physics, Graduate School of Advanced Science and Engineering, Faculty of Science and Engineering, Waseda University, 3-4-1 Okubo, Shinjuku, Tokyo 169-8555, Japan}
\affiliation{Waseda Research Institute of Science and Engineering, Faculty of Science and Engineering, Waseda University, 3-4-1, Okubo, Shinjuku, Tokyo 169-8555, Japan}

\author{Satoshi Yamanaka}
\affiliation{General Education Department, National Institute of Technology, Toba College, 1-1, Ikegami-cho, Toba, Mie 517-8501, Japan}
\affiliation{Research Center for Space and Cosmic Evolution, Ehime University, 2-5, Bunkyo-cho, Matsuyama, Ehime 790-8577, Japan}
\affiliation{Waseda Research Institute of Science and Engineering, Faculty of Science and Engineering, Waseda University, 3-4-1, Okubo, Shinjuku, Tokyo 169-8555, Japan}

\author{Toru Yamada}
\affiliation{ISAS/JAXA}


\begin{abstract}
The morphological study is crucial to investigate the connections between active galactic nuclei (AGN) activities and the evolution of galaxies. Substantial studies have found that radiative-mode AGNs primarily reside in disk galaxies, questioning the merger-driven mechanism of AGN activities. In this study, through S{\'e}rsic profile fitting and non-parametric morphological parameter measurements, we investigated the morphology of host galaxies of 485 optical variability-selected low luminosity AGNs at $z\lesssim4.26$ in the COSMOS field. We analyzed high-resolution images of the Hubble Space Telescope to measure these morphological parameters. We only successfully measured the morphological parameters for 76 objects and most AGN hosts ($\sim70\%$) were visually compact point-like sources. We examined the obtained morphological information as a function of redshift and compared them with literature data. We found that these AGN host galaxies showed no clear morphological preference. However, the merger rate increased with the higher hosts' SFRs and AGN luminosity. Interestingly, we found ongoing star formation consistent with the typical star forming populations in both elliptical and spiral galaxies while these two types of galaxies were more symmetric than normal star forming galaxies. These results suggested that optical variability-selected AGNs \textcolor{black}{have} higher probabilities to reside in elliptical galaxies than infrared-selected AGNs (IR-AGNs), whose host galaxies had a strong disk-dominance, and supported recent studies that the AGN feedback
could enhance star forming activities in host galaxies.
\end{abstract}

\section{Introduction} \label{sec:intro}
Galaxy morphology is a direct means that reveals the interaction of galaxies with their environments and the impact of the internal perturbation.
Based on galaxies' visual appearance, the morphological study provides a unique method to investigate the evolution of galaxies.
In 1926, Edwin Hubble proposed a galaxy morphology classification scheme, the so-called Hubble sequence, which divided galaxies into four classes: ellipticals, lenticulars, spirals, and irregulars \citep{1926ApJ....64..321H}.
In addition, in 1959, de Vaucouleurs extended Hubble's scheme, taking rings into consideration \citep{1959HDP....53..275D}.
Recently, a well-known project, Galaxy Zoo, proposed more detailed morphological classifications, including disturbed features and cigar-shape classifications for edge-on galaxies \citep{2008MNRAS.389.1179L, 2013MNRAS.435.2835W}.
Along with purely visual investigation, parametric (S\'ersic index $n$) and non-parametric methods (Gini $(G)$, $M_{20}$, Concentration $(C)$, \textcolor{black}{Asymmetry} $(A)$, \textcolor{black}{Smoothness} $(S)$, and \textcolor{black}{Ellipticity}) have been proposed to quantitatively describe the light distribution within a galaxy. 
Galaxies difficult to visually classify at high-redshifts or due to edge-on structures can be more accurately distinguished through these parameters.
For the S\'ersic index, \citet{2004ApJ...604L...9R} found $n=2$ efficient enough to separate early- and late-type galaxies and \citet{2011ApJ...743...96C} successfully applied this value to high-$z$ HST galaxies. 
\citet{2003ApJS..147....1C} presented the values of $G$, $A$, and $S$ for nearby galaxies and $G$ decreased as the galaxies varied from ellipticals to disks, and irregulars, whereas $A$ and $S$ increased.

Blackholes (BH) are ubiquitously found in the centers of galaxies. 
The co-evolution of a galaxy and its central BH is one of the most attractive issues in modern astronomy.
Unfortunately, owing to the limitation of observational techniques, we can only investigate such connections via a proxy, which is known as the active galactic nucleus (AGN). 
In most cases, an AGN resides in the center of its host galaxy on a very small scale and is fueled by the gas inflow--an accretion onto the BH. 
Generally, AGNs that have dust-obscured structures, along with broad/narrow line regions (BLRs/NLRs), are rich in emission lines.
These AGNs are called radiative-mode.
They are more likely to be held by moderately massive ($M_\star\sim 10^{10}\rm{\ \textcolor{black}{to}\ a\ few\ times} \times10^{11} M_\odot$) disk systems undergoing star forming activities with SFRs corresponding to those of the typical star forming populations at epochs of up to $z\sim2$ \citep{2012ApJ...757...81B, 2012ApJ...760L..15H}. 
On the other hand, jet-mode AGNs that are radiatively inefficient with advection-dominated accretion flows (ADAF) tend to reside in massive ellipticals, or spheroid systems \citep{2014ARA&A..52..589H}.
Such co-evolution studies help us understand the growth mechanism of BHs, how AGNs influence the star formation rate (SFR), and the global structure of the host galaxy.

Before the detailed studies is the selection of AGNs out of astronomical objects such as normal galaxies.
One primary approach to select AGNs is based on X-ray observations. \citet{2011A&A...535A..80M} selected 142 Type-2 QSOs in the COSMOS field via XMM-Newton observatory at $L_X\mathrm{[0.5-10\ keV]=10^{44}-10^{45}\ erg\ s^{-1}}$ and $\sim 0.8<z<2.0$. 
They found the majority of their objects residing in early-type galaxies, and the minority belongs to prominent disk systems or mergers.
Meanwhile, at $z\sim1$, about $62\pm7\%$ Type-2 QSO hosts are actively forming stars and this tendency becomes more apparent at higher redshifts, suggesting an evolutionary effect.
The evolution of the specific SFR (sSFR) of these QSO hosts along the redshift excellently agrees with that of normal star forming galaxies (SFGs) at $1<z<3$.
In addition, using $Herschel$ PACS observations, \citet{2012A&A...540A.109S} found the hosts of their X-ray-selected radiative-mode AGNs in GOODS-S and -N fields at $\sim0.5<z<2.5$ exhibited enhanced SFRs and sSFRs in comparison with mass-matched inactive galaxies, indicating that SFGs are more likely to host AGNs.

Infrared (IR) observations are paramount for explorations on the obscured AGNs and act as complements to X-ray data. 
This method can unveil AGNs that even the deepest X-ray observations cannot detect. 
\citet{2012MNRAS.425L..61S} studied the nature of quasar host galaxies based on mid-IR (MIR, $24\mu m$) selected dust-obscured galaxies (DOGs) at $z\sim2$ using Hubble Space Telescope WFC3/IR imaging data. 
Contrary to X-ray-selected Type-2 AGN hosts analyzed by  \cite{2011A&A...535A..80M}, most of their DOGs are disk systems. 
Further, they argued that only a minority of these objects are mergers. 
This merger rate of AGN hosts is also supported by the study on IR-AGN hosts up to $z\sim2.5$ \citep{2017ApJS..233...19C}; they suggested that the merger rate of most luminous AGNs with $\log{(L_{IR}/L_\odot)}\sim12.5$ could be as large as 50\%.\par

According to the sample selections, redshifts, and techniques used for the studies, the relations between the host galaxy's morphology and AGN activities are still under debate. 
\citet{2007ApJ...660L..19P} selected 94 AGNs based on X-ray observations at $0.2<z<1.2$ and found most of their host galaxies classified as E/S0/Sa; non-parametric techniques revealed no evident morphological preference in the hosts of IR-AGNs. 
\citet{2009ApJ...691..705G} performed a two-component decomposition for $\sim400$ X-ray-selected AGN hosts at $0.3<z<1.0$ and found that the morphology of host galaxies spanned a wide range from bulge-dominated ($1.5<$ S\'ersic index $n<10$) to disk-dominated ($n\sim1$) systems and peaks between these two systems. \citet{2014MNRAS.439.3342V} simulated AGN components and co-added them with stellar mass-matched control samples at $0.2<z<0.8$, finding similar distributions of asymmetries, S\'ersic indices, and ellipticities between AGN hosts and control sample galaxies. 
At $1.25<z<2.67$, \citet{2011ApJ...743L..37S} considered point source components and suggested that over half of their 57 X-ray AGN hosts resided in disk-dominated systems.

Low luminosity AGNs (LLAGNs) suffer contaminations by light from their host galaxies, where usual color selection techniques do no work. 
Therefore, the flux variability, possibly arising from the instability of the accretion disk, in multiepoch observations can be employed as a new selection technique.
By studying the host galaxies of MIR variability-selected AGNs, \citet{2012MNRAS.426..360V} found common disturbed morphological features, and that it was secular processes--tidal processes and minor mergers in particular--that triggered LLAGN activities rather than major mergers.

In this study, we explore the morphology of host galaxies whose central AGNs are selected based on their optical variabilities up to $z\sim3$. 
Such a method provides a more efficient selection of unobscured AGNs with low bolometric luminosities compared with X-ray or IR observations.
Accompanied by this low obscuration is the high contamination of their light to the host galaxies. 
Further, high redshifts beget a cosmological surface brightness dimming of the galaxies, making it more difficult to visually determine the morphological classification. 
Considering these, we employ a 2D surface brightness fitting (S\'ersic index), which includes a correction for central brightness excess, and non-parametric methods less affected by an AGN component but also corrected. 
To obtain detailed information about hosts' SFRs, masses, and AGN fractions to investigate AGN-host connections, the spectral energy distribution (SED) fitting is performed using IR to X-ray photometric data.

The rest of this article is organized as follows: In Section \ref{sec:sample}, we describe the samples as well as imaging and photometric data. Section \ref{sec:methods} introduces both parametric and non-parametric methods, along with how we perform the SED fitting. 
The results of morphological measurements are presented in Section \ref{sec:results}. 
Indirect results derived from different methods, as well as implications are presented in Section \ref{sec:discus}.
Finally, we summarize in Section \ref{sec:conc}. We use the following $\mathrm{\Lambda CDM}$ cosmological parameters to calculate the radius in the unit of kpc: $\mathrm{H_0=70\ km\ s^{-1} Mpc^{-1}},\ \Omega_M=0.27,\ \Omega_\Lambda=0.73$.
All magnitudes in this article follow the AB system \citep{1983ApJ...266..713O}.

\section{Sample and Data}\label{sec:sample}
\subsection{Optical Variability-Selected AGNs}
The parent catalog of the AGNs examined in this study is based on optical variability-selection (Subaru Hyper-Suprime-Cam $g, r, i, z$, \citealt{2018PASJ...70S...3F,2018PASJ...70...66K,2018PASJ...70S...2K,2018PASJ...70S...1M}) by \cite{2020ApJ...894...24K}, consisting of 491 variability-selected AGNs in the Cosmic Evolution Survey (COSMOS, \citealt{2007ApJS..172....1S}) field. 
The brightness of the AGNs in this catalog ranges from 17.56 to 25.87 in \textit{i}-band magnitude, and the redshifts are spanned up to $z=4.26$.

\begin{figure}
\begin{center}
\includegraphics[width=0.47\textwidth]{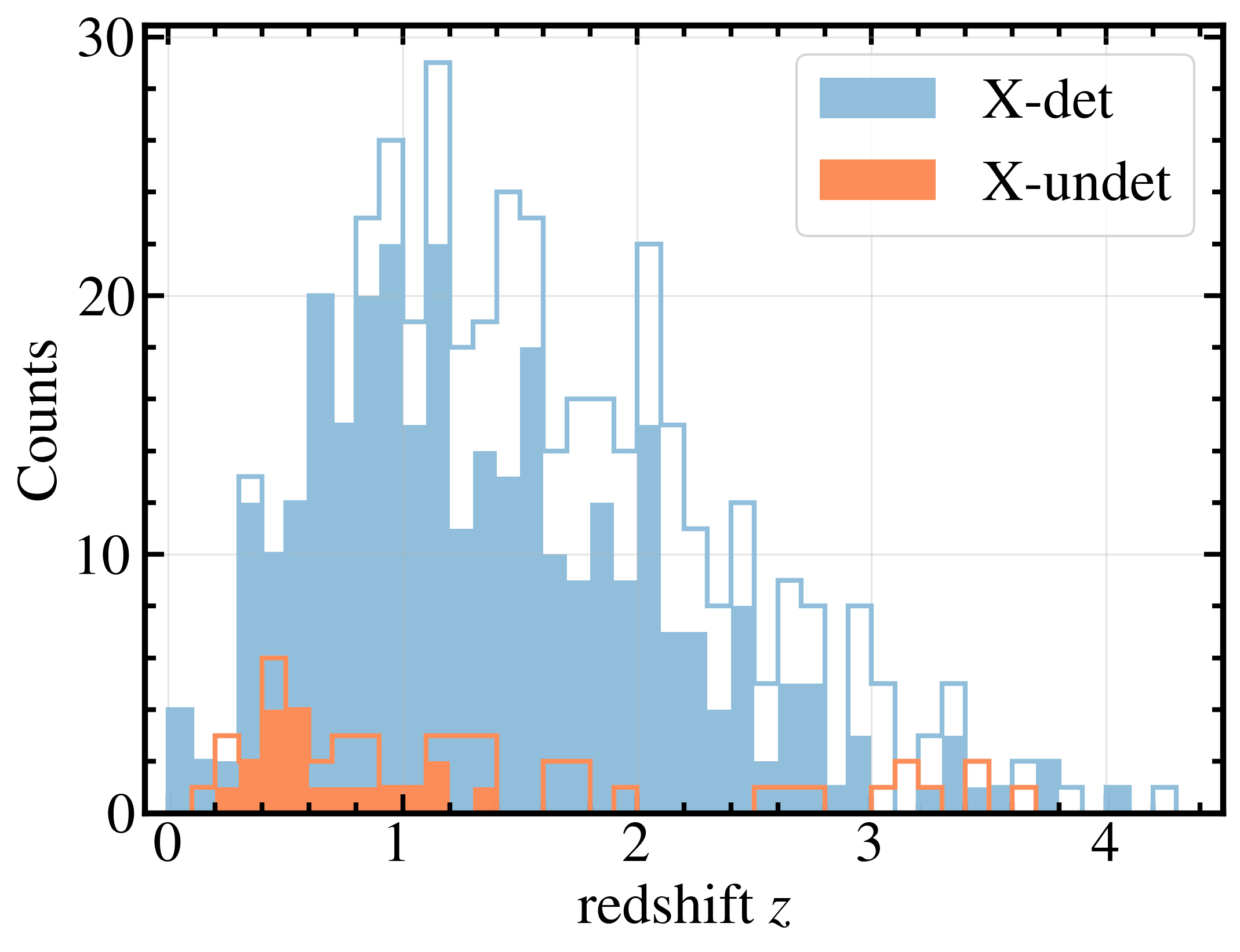}
\caption{Redshift distribution of the optical variability-selected AGNs. The blue and orange histogram represents X-ray detected and undetected AGNs, respectively. The filled and open histogram displays the number of objects with spectroscopic redshift and photometric redshift in each bin, respectively.\label{fig:1}}
\end{center}
\end{figure}

The redshift distribution of these variability-selected AGNs is plotted in Fig.~\ref{fig:1}. 
Of all these objects, 441 ($\sim90\%$) were detected in X-ray observations with Chandra, and 337 objects ($\sim69\%$) had spectroscopic redshifts. 
The available spectroscopic redshifts are provided by $z_{spec}$ in the Subaru Hyper-Suprime-Cam (HSC) PDR2 catalog \footnote{\textcolor{black}{https://hsc-release.mtk.nao.ac.jp/doc/index.php/database-2/}}, which are collected from zCOSMOS DR3 \citep{2009ApJS..184..218L}, PRIMUS DR1 \citep{2011ApJ...741....8C, 2013ApJ...767..118C}, VVDS \citep{2013A&A...559A..14L}, SDSS DR12 \citep{2015ApJS..219...12A}, FMOS-COSMOS \citep{2015ApJS..220...12S}, 3D-HST \citep{2016ApJS..225...27M} and DEIMOS 10 Spectroscopic Survey Catalog (DEIMOS catalog, \citealt{2018ApJ...858...77H}). 
If no spectroscopic redshift is available, \textcolor{black}{$\rm{z\_best}$} (best photometric redshift) in the Chandra catalog \citep{2016ApJ...817...34M} is used for X-ray-detected objects, and \textcolor{black}{$\rm{z\_{PDF}}$} (photometric redshift measured using the galaxy templates) in COSMOS2015 catalog \citep{2016ApJS..224...24L} is used for X-ray-undetected objects.

\subsection{Imaging Data for Morphological Classification} \label{sec:image}
To visually inspect the morphology of host galaxies of the optical variability-selected AGNs, we choose the $I$-band ($F814W$) imaging data obtained with the Hubble Space Telescope/Advanced Camera for Surveys (HST/ACS) from the COSMOS-HST Treasury project \citep{2007ApJS..172..196K,2010MNRAS.401..371M}.
We searched around the coordinate of each object recorded in the Subaru HSC catalog offered by \citet{2020ApJ...894...24K} within a radius of $1.5''$ in the COSMOS-HST catalog to obtain the HST cutout for each AGN host galaxy, and 485 of 491 objects were found in the COSMOS-HST Treasury project \citep{2007ApJS..172..196K,2010MNRAS.401..371M}.
The cutout of each object had 202 pixels in height and width with the resolution per pixel of $0.03''$, corresponding to $6''\times6''$ in size. 

Checking the imaging data, we found that many host galaxies were contaminated by their central AGNs, leading to bright point spread function (PSF)-like components in their centers, or completely dominated by the PSF. 
Therefore, we considered the PSF component a necessary term to fit the 2D S\'ersic brightness profile. 
 
We used the Tiny Tim HST PSF Modeling Tool \citep{2011SPIE.8127E..0JK} to generate the PSF of the $I$-band ($F814W$) images. 
The final modeled PSF was the average of the stacking of 17 PSFs modeled with 17 types (O, B, A, F, G, K, and M, including intermediate types) of stars, which were provided by Tiny Tim. 
The PSFs modeled with some types of stars showed unexpected gaps or bumps in their radial light profiles. 
We performed the stacking to avoid the possible influences of such gaps on the successive measurements.

\begin{figure}
\begin{center}
\includegraphics[width=0.47\textwidth]{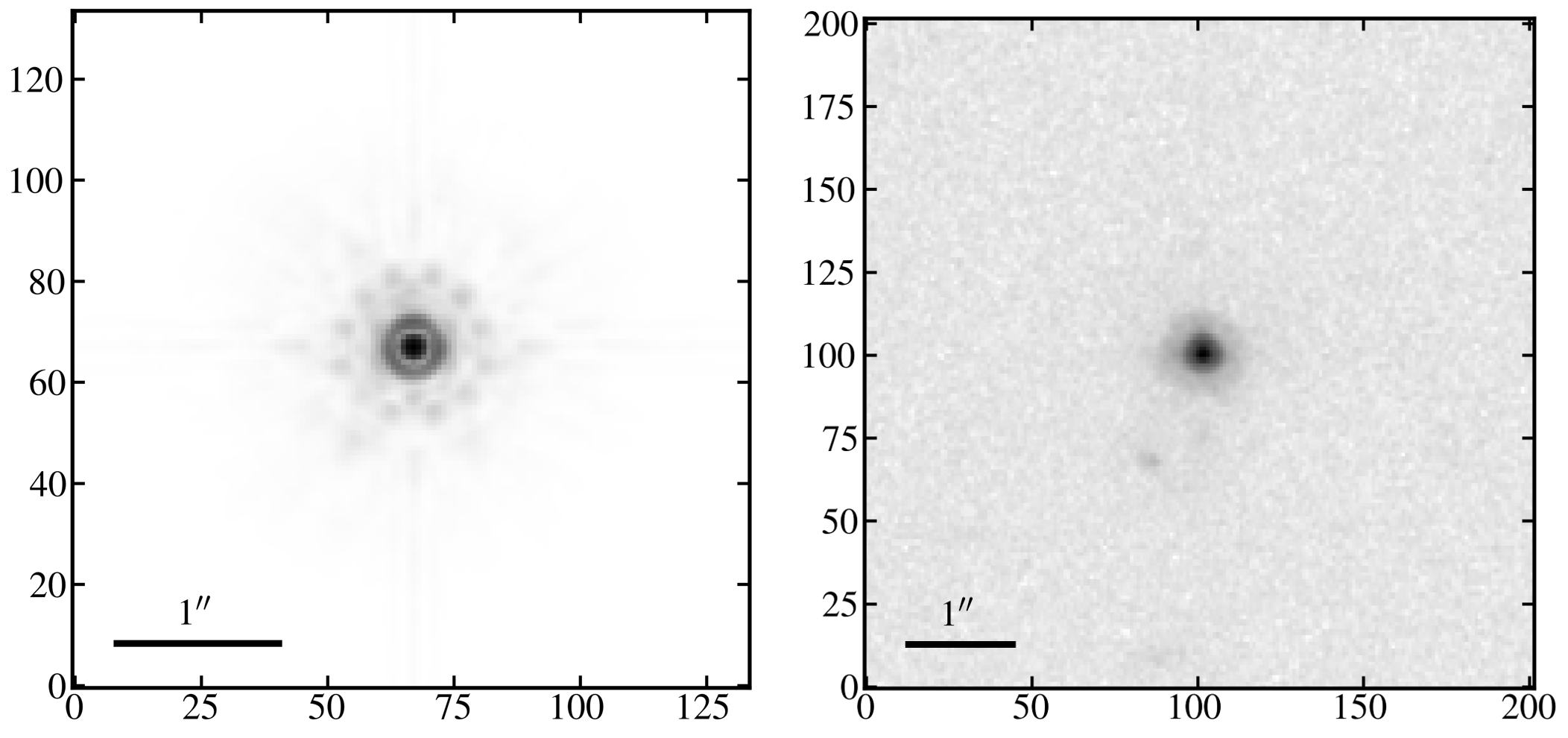}
\caption{Left panel: PSF modeled by Tiny Tim Modeling Tool \citep{2011SPIE.8127E..0JK} with $\mathrm{FWHM}=2.853$ pixels. 
Right panel: ID002, an object at $z=3.599$ with $\mathrm{FWHM}=2.75$ pixels, completely convolved to be a PSF-dominated object due to its extremely compact size. 
The small difference of 0.1 pixels between the two FWHMs is believed to be acceptable and sufficient for the successive morphological analysis. \label{fig:psf}}
\end{center}
\end{figure}

A real point source convolved with the real PSF should always be larger than the size of the PSF.
To ensure the reliability of the modeled PSF, its full-width at half-maximum (FWHM) was measured via $\texttt{IRAF-psfmeasure}$ command, and compared with a few PSF-dominated objects (the entire object is visually in shape of the PSF), which could be considered real PSFs.
The modeled one had an FWHM of 2.853 pixels, whereas two objects at $z>4$ both had FWHM larger than 3 pixels, and an object at $z=3.599$ had $\mathrm{FWHM}=2.75$ pixels.
Although the FWHM of this object at $z=3.599$ was smaller than that of the modeled PSF, we considered that such a 0.1-pixel difference was insignificant. 
Thus, we believe this modeling is sufficient for our purpose of involving it as a necessary term for the 2D S\'ersic fitting. 
Fig.~\ref{fig:psf} depicts our modeled PSF and the $z=3.599$ object as an example of PSF-dominated host galaxies.

\subsection{Photometric Data} \label{sec:sedata}

We obtained multi-wavelength photometry from the COSMOS2015 catalog \citep{2016ApJS..224...24L}. 
The data include near-ultraviolet (NUV, 200-300 nm) and far-UV (FUV, 100-200 nm) observations from GALEX (The Galaxy Evolution Explorer satellite, \citealt{2007ApJS..172..468Z,2007ApJS..172...99C}), UV observations (300-400 nm) from the Canada-France Hawaii Telescope (u-band, CFHT/MegaCam), and optical data from COSMOS-20 survey taken with Subaru Suprime-Cam in five broad bands (B, V, i, r, z), six intermediate bands (IB427, IB464, IB505, IB574, IB709, and IB827), and two narrow bands (NB711 and NB816) \citep{2007ApJS..172....9T,2015PASJ...67..104T}. 
In the near-IR (NIR) range, Y-, J-, H-, Ks-band data taken with WIRCam and Ultra-VISTA \citep{2010ApJ...708..202M, 2012A&A...544A.156M} were used.
In the MIR range, the SPLASH-COSMOS survey (Spitzer Large Area Survey with HSC; PI: P. Capak), S-COSMOS  \citep{2007ApJS..172...86S}, and the Spitzer Extended Mission Deep Survey \citep{2013ApJ...769...80A} provide data in [3.6], [4.5], [5.8], and [8.0] $\mu$m. 
The units of these photometry were converted from the Absolute Magnitude (AB mag) to flux densities ($F_\nu$) with the unit mJy ($F_\nu=10^{(8.90-m)/2.5}*1000$). 
The fractional errors in AB mag were used as the uncertainties on $F_\nu$.

Many objects were detected in X-ray observations, and we selected Chandra 0.5-10 keV band data originally from the "Chandra COSMOS Legacy" survey \citep{2016ApJ...819...62C} for our X-ray-detected objects and converted their flux unit of $\mathrm{erg\ s^{-1}\ cm^{-2}}$ to the unit of flux density mJy (corresponding to the frequency of $2.297\times10^{15} \mathrm{Hz}$). 
We assumed a 10\% fractional difference of the observed flux as the uncertainty, which was typical for our objects based on the catalog by \citet{2016ApJ...819...62C}.

\section{Methods} \label{sec:methods}
\subsection{Parametric Method} \label{sec:2dfitting}
The basic parametric method includes the measurement of the S\'ersic index ($n$) and the effective radius ($r_{\rm e}$, the radius that encloses 50\% of the galaxy's total flux).
We performed 2D S\'ersic+PSF profile fitting by employing $\texttt{Astropy}$ \citep{astropy:2013, astropy:2018} and its affiliated package $\texttt{statmorph}$ \citep{2019MNRAS.483.4140R}.
This package was also used for the non-parametric measurements in the next subsection.

To perform the fitting, the segmentation map, a 2D array with the same size as the cutout, that labels the pixel belonging to the source should be created in advance.
We used $\texttt{photutils}$ \citep{larry_bradley_2020_4044744} to create this image segmentation.
For the first step, a smoothing box, which was defined as a 2D circular Gaussian kernel with an FWHM of 3 pixels, was used to smooth the cutout image and increase the object's detectability. 
Next, we estimated the background noise of the cutout and detected the target object according to the threshold defined as $2\sigma$ per pixel above the smoothed background noise.
The segmentation maps were generated based on this threshold and the FWHM size of the smoothing box.
For several objects too faint or showing irregular structures, it might fail in basic morphological measurements or 2D S\'ersic fitting.
To correctly perform the fitting we adjusted the threshold ($1.5-3\sigma$ range) and the FWHM size of the smoothing box. 
We compared the measured values before and after adjusting the parameters for some successfully measured objects and found only insignificant changes on both parametric and non-parametric measurements. 

As described in \S\ref{sec:image}, an AGN and its host galaxy was a composite of two components, which meant we needed a decomposition to perform a more accurate fitting. 
Otherwise, the bright AGN in the center will enhance the central brightness, resulting in biased S\'ersic indices and non-parametric parameters (see \S\ref{sec:correction}).
Although $\texttt{statmorph}$ could perform a de-convolution of the galaxy profile convolved with the PSF, it could not decompose the host galaxy and its central AGN. 
We describe how we correct this systematic error in \S\ref{sec:correction}\par 

\begin{figure*}[ht]
\begin{center}
\includegraphics[width=\textwidth]{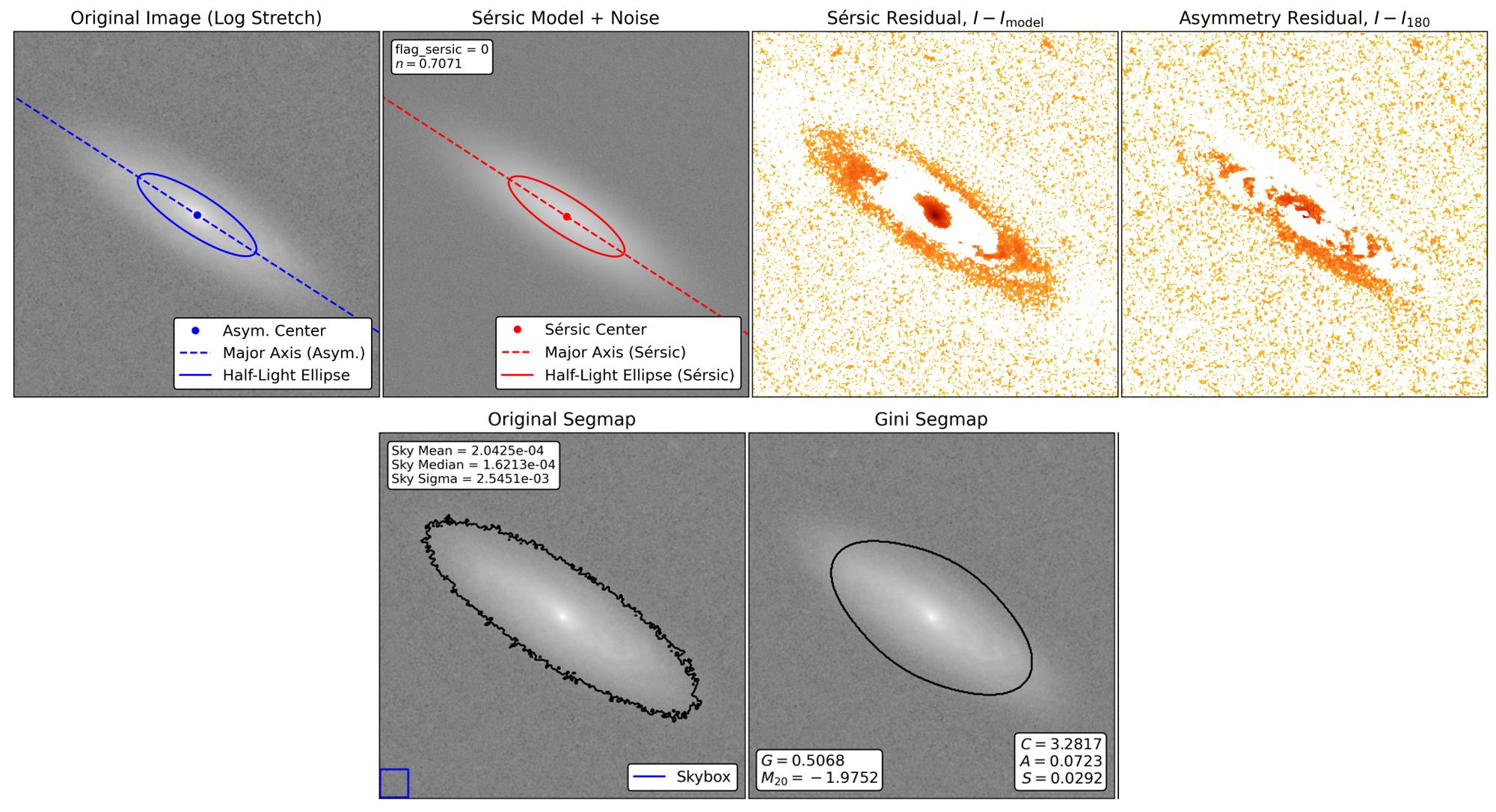}
\caption{Output plot of the object ID190, a spiral galaxy at $z=0.184$. $\texttt{Original Image}$ shows the cutout of the original galaxy image. The blue dot represents the center of the galaxy, whereas the blue dashed and solid lines indicate the orientation and extent of the elliptical aperture that encloses half of the total flux of the galaxy.
$\texttt{S\'ersic model + Noise}$ shows the fitted S\'ersic model with background noise added. 
On the left-hand box, $\texttt{flag\_sersic==0}$ indicates a successful 2D S\'ersic fitting and $n$ is the value of measured S\'ersic index. The basic shape and size measurements derived from the S\'ersic fitting are indicated by the red point and lines, respectively.
$\texttt{S\'ersic Residual}$ shows the residual of the original galaxy after subtracting the S\'ersic fitting.
$\texttt{Asymmetry Residual}$ shows the residual of the original galaxy subtracted by half of the image rotated by 180 degrees.
$\texttt{Original Segmap}$ gives the segmentation map created by $\texttt{photutils}$ and values of the background noise. 
The detection thresholds and sizes of smoothing boxes were adjusted to ensure an enclosure of the entire galaxy including substructures as much as possible.
The non-parametric parameters $(CMCAS)$ are shown in $\texttt{Gini Segmap}$, wherein the region enclosed by the black solid line is the $\texttt{Gini\_segmentation}$ defined by the Petrosian radius ($\eta=0.2$).
\label{fig:statmorph}}
\end{center}
\end{figure*}

\subsection{Non-parametric Method} \label{sec:gmca}
$\texttt{statmorph}$ also provides a set of non-parametric parameter measurements \citep{2004AJ....128..163L, 2003ApJS..147....1C}. 
This method was used to measure the light distribution of a galaxy. 
Non-parametric measurements performed in this study include five parameters as described below.

The first is Gini ($G$) defined by the Lorentz curve of the galaxy's light distribution. 
This parameter describes whether the light evenly distributes among all pixels or concentrates on several pixels.
The value of $G$ ranges from 0 to 1; 1 indicates that a single pixel encloses all brightness and 0 indicates all pixels have the same brightness.
Notably, $G$ does not consider spatial positions. 

The second parameter is $M_{20}$, the second order moment of the galaxy's brightest region enclosing 20\% of the total flux of the galaxy \citep{2004AJ....128..163L}. 
It is a tracer of any bright nucleus, bar, spiral arm, and off-center star-cluster within a galaxy. 
The value of $M_{20}$ is always negative; the lower it is, the higher the concentration anywhere within the galaxy.

The third parameter is Concentration ($C$), simply defined as follows \citep{2003ApJS..147....1C}:
\begin{equation}
    C=5\times log\left(\frac{r_{80}}{r_{20}}\right),
\end{equation}
where $r_{80}$ and $r_{20}$ are radii enclosing 80\% and 20\% of the galaxy's light, respectively. 
Concentration is directly related to the galaxy's central brightness; a larger value indicates a brighter central region and larger S\'ersic index.

The fourth parameter is Asymmetry ($A$), which is an indicator of whether there exist non-symmetric components within the galaxy. 
Spiral galaxies usually have large Asymmetry due to their spiral arms. 
Besides, in-homogeneous star forming activities will also result in asymmetric structures. 
Although the value of $A$ is usually positive, since such measurement depends on background noise, negative values may appear in low signal-to-noise ratio (SNR) cases.

The last parameter is Smoothness ($S$, or Clumpiness), which describes if there is any clumpy structure within the galaxy.
Elliptical galaxies are typical smooth systems holding small values of Smoothness, and star forming galaxies tend to have more clumpy regions.

To calculate these five parameters, $\texttt{statmorph}$ would create a $\texttt{Gini\_segmentation}$ based on the Petrosian radius ($r_{petro}$, \citealt{1976ApJ...209L...1P}), which is given by
\begin{equation}
    \eta(R)=\frac{I(R)}{\langle I(<R) \rangle},
\end{equation}
where $I(R)$ is the surface brightness at a radius $R$, and $\langle I(<R) \rangle$ is the mean surface brightness within the radius $R$. 
We choose $\eta(R=r_{petro})=0.2$ \citep{2004AJ....128..163L} to perform calculations. 
In fact, we tested measurements with different values for $\eta$ and found that the $\texttt{Gini\_segmentation}$ would ignore substructures, such as spiral arms, as $\eta$ increased and the measured values only varied slightly. 
Fig.~\ref{fig:statmorph} gives an example output plot of the measurement, including S\'ersic fitting and basic parameter measurements.

After checking the results, we found that over $70\%$ of our objects had $S=0$. 
We investigated the outputs and found all these objects were PSF-like, indicating a significant optically-unobscured AGN fraction. 
For any of these objects, there was no reliable $\texttt{Gini\_segmentation}$ as well because the $r_{petro}$ always merely enclosed the bright core (with almost the FWHM size) of the PSF-like object, making it impossible to measure the non-parametric morphological parameters of the host galaxies. 
Then, we further checked the $\texttt{Gini\_segmentation}$ of the remaining objects so that the non-parametric parameters can be measured on the basis of a reliable enclosure of the galaxy's light.
As a result, we found many cases where the objects had no reliable $\texttt{Geni\_segmentation}$ with non-parametric parameters concentrating around the combination of $G=0.5\pm0.02$, $M_{20}\sim-1.74$, and $C=3.0\pm0.2$.
In $G-M_{20}$ diagnostics, these objects were all classified as Sb/Sc/Irr, leading to an overestimation of the fraction of disk systems (see \S\ref{sec:diag}).
Real objects with this combination of $G$, $M_{20}$, and $C$ were also excluded.

\subsection{Correction for Effects of the Central PSF Components} \label{sec:correction}

For the remaining objects, there were central bright PSF-like components in many cases.
To correct any systematic effect due to the PSF-like components, in particular, to investigate how the S\'ersic index was affected, we modeled AGN host galaxies.
The modeling includes several preset parameters: S\'ersic index ($n$), effective radius ($r_{\rm e}$), light intensity $I_{\rm e}$ at the effective radius, ellipticity, and ratio of the total flux of the host galaxy and AGN ($f_{\rm host}:f_{\rm AGN}$). 

As the first step, we modeled 300 galaxies whose intrinsic S\'ersic index ranged from 0.1 to 5.0 (50 samples) and ellipticities ranged from 0 to 0.5 (six groups), both with a 0.1 increase between two consecutive setups.
Then, we considered three different effective radii ($r_{\rm e}=5, 10, 15$ pixels) and obtained $300\times3=900$ galaxies.
The intensity at the effective radius was assumed to be $I_{\rm e}=10$ electrons s$^{-1}$ pix$^{-1}$, which was arbitrarily chosen, but scaled by an SNR, as described below.
The unit is the same as the actual HST cutout images.
To model an AGN component, the brightest pixel of the modeled galaxy made in the first step was co-added with the value of the total flux multiplied by a factor.
This factor was assumed to be $f_{\rm AGN}$, given by the following combinations:  
$f_{\rm host}:f_{\rm AGN}=$ 1:0.01, 1:0.1, 1:0.2, ..., 1:0.9, 1:1.0, 1:1.5, 1:2, 1:2.5, 1:3, 1:4, 1:5.
Hence we had $900\times17=15300$ galaxies harbouring AGNs modeled in total.
These model galaxies, including the AGN component, were convolved with the modeled $F814W$ PSF.
Then, we added noise in each pixel of the images following an error function whose standard deviation $\sigma=0.4$ electrons s$^{-1}$ pix$^{-1}$, corresponding to an $\mathrm{S/N}=25$ at the effective radius. 
The SNR at the effective radius was chosen as a typical value observed in our actual AGN host galaxies.\footnote{In this procedure, the total flux of model galaxies without the AGN component, $f_{\rm host}$, was calculated by the method as follows.
We convolved the model galaxies with the $F814W$ PSF and co-added with the pixel noise following an error function with $\sigma=0.1$ electrons s$^{-1}$ pix$^{-1}$.
These high $\mathrm{S/N}=100$ model galaxies were used only to make a reliable segmentation map, wherein we calculated the total flux, $f_{\rm host}$ as the sum of the pixel counts.}

By comparing the assumed values of S\'ersic index and $r_{\rm e}$ with the values measured by $\texttt{statmorph}$, we found many cases where the differences were unacceptably large (measured $n>10$) and the host galaxy was heavily contaminated by the AGN.
Too make these models good matches of real objects, we adapted the following criteria to exclude invalid results:

\begin{enumerate}\label{crit}
\item Either $\texttt{flag\_sersic}==0$ or $\texttt{flag}==0$ (any error flag returned by $\texttt{statmorph}$).
\item Either $r_{\rm e}$ or $sersic\_rhalf$\footnote{$r_{\rm e}$ indicates the effective radius of the object in the original image, and $sersic\_rhalf$ indicates the effective radius of the model fitted by 2D S\'ersic fitting} is smaller than that of the $F814W$ PSF (2.19 and 1.52 pixels, respectively).
\item The size of the  $\texttt{Gini\_segmentation}$ is smaller than the effective radius as well as 50\% of the size of the original segmentation.
\item The measured $Smoothness$ is equal to 0.
\item Cases without reliable $\texttt{Gini\_segmentation}$, for which $G=0.5\pm0.02$, $M_{20}\sim-1.74$, and $C=3.0\pm0.2$.
\item Objects with failure in the calculation of the total flux, because of which an AGN component cannot be modeled.
\end{enumerate}
% \vspace{-0.5cm}

By adapting these criteria, 7,697 modeled AGN hosts were excluded and there were 7,603 successfully modeled AGN host galaxies.
Except for the last one, the other criteria were applied to real objects to exclude unsuccessful measurements, ensuring a fair comparison between real and simulated AGN$+$host galaxies.
\textcolor{black}{As mentioned in \S\ref{sec:gmca} that over 70\% objects had $S=0$, the fourth criterion produced the greatest rejection of the valid objects.
These host galaxies suffered both cosmological surface brightness dimming and heavy contamination from their central AGNs, resulting in a complete PSF-like morphology, of which object ID002 in the right panel of Fig.~\ref{fig:psf} is representative.}
The correction method for the S\'ersic index is described in Appendix~\ref{apd} in detail.
In addition, if the fraction of $r_{\rm e}$ to $sersic\_rhalf$ of the real object is out of the range of the modeled hosts within the corresponding effective radius, there is no available correction for the object, and it was also excluded (see Appendix \ref{apd}).

Corrections for $G$, $M_{20}$, and $C$ were also made through simulations.
Unlike S\'ersic index that depends on a certain mathematical form, non-parametric parameters are much less affected by an AGN component only if $\texttt{Gini\_segmentation}$ has a reliable enclosure of the host galaxy's light and the fractional differences of values ($\mathrm{(value_{galaxy+AGN}- value_{galaxy})/value_{galaxy+AGN}}$) for a galaxy with and without an AGN can be applied for the correction.

The basic parameters of the simulation are the same as those for the correction for S\'ersic index other than the amplitude of the point source (or the total flux of the AGN) and the definition of $f_{\rm host}:f_{\rm AGN}$.
Apart from the simulation parameters, the non-parametric parameters were measured for the host galaxies without AGN components and then measured after the two components were co-added so that the fractional differences could be calculated.
In this simulation, the total flux is set to 10 and $f_{\rm host}$ and $f_{\rm AGN}$ were the pixel counts of the center-one pixel (which is the brightest one) of the galaxy and the AGN, respectively.
Before co-adding the two components, both of them were convolved with the PSF and then normalized so that the brightest pixels had pixel counts equal to 1; thus, the $f_{\rm host}:f_{\rm AGN}$ could be well controlled.
There were four groups for the flux ratio between the host and AGN: $f_{\rm host}:f_{\rm AGN}=3:1, 1:1, 1:4, 1:6$. 
For $f_{\rm host}:f_{\rm AGN}=3:1$, there was no visible PSF, and the other three had visible PSF components.
Thus, the modeled host galaxies were further divided into two groups: with or without visible AGN components.
Modeled host galaxies without visible AGN components corresponded to real objects whose central AGNs could not be visually confirmed or those with $r_{\rm e}>16$ pixels, whereas modeled hosts with visible AGN components were matched to real objects that had apparent AGN components.
The fractional differences are listed in Table \ref{tab:corrfac}. 
To correct the measurement biases, the following equation is adapted:
\begin{equation}
    NP_\text{corrected}=NP_\text{measured}*(1-f_{NP}),
\end{equation}
where $NP$ represents the non-parametric parameter, and $f_{NP}$ is the fractional difference of the corresponding parameter.

Through modeling, it is found that galaxies with small $r_{\rm e}$ and large S\'ersic index ($n\geq3$) are convolved to be the PSF-shape even if they do not have AGN components. 
By applying the criteria, along with those failing in the measurement, 349 objects were excluded, including many spheroid systems ($n>2$) before they could be further discussed.
This might result in potential biased dominant morphology of the whole sample with 485 AGN host galaxies.

\begin{deluxetable}{cccc}[htbp]\label{tab:corrfac}
\tablecaption{Fractional differences of $G$, $M_{20}$, and $C$ measurements due to the presence of an AGN. }
\tablehead{ & & \colhead{$r_{\rm e}$$^a$} & \\
\cline{2-4}
\colhead{Criterion} & \colhead{5} & \colhead{10} & \colhead{15} }
\startdata
\multicolumn{4}{c}{Gini: $G$} \\
PSF visible & 0.059 & 0.087 & 0.067\\
PSF not visible & 0.016 & 0.013 & 0.008\\
\hline
\multicolumn{4}{c}{$M_{20}$} \\
PSF visible & 0.078 & 0.111 & 0.124\\
PSF not visible & 0.026 & 0.025 & 0.020\\
\hline
\multicolumn{4}{c}{Concentration: $C$} \\
PSF visible & 0.115 & 0.183 & 0.124\\
PSF not visible & 0.040 & 0.024 & 0.019\\
\enddata
\tablecomments{$^a$ The effective radii are in unit of pixel with a resolution of $0.03''$/pixel. \label{tab:GMC}}
\end{deluxetable}
% \vspace{-1cm}

\begin{deluxetable*}{lll}[htbp!]
\tabletypesize{\small}
\tablecaption{X-CIGALE user-specified components for SED fitting described in \ref{sec:xcigale}}\label{tab:2}
\tablehead{Parameter & Values & Description}
\startdata
\multicolumn{3}{c}{Delayed star formation history: $\texttt{delayedSFH}$ } \\
$\tau_\mathrm{main}$ & 300, 500, 800, 1000, 1500, 2000 & e-folding time of the main stellar population model \\*  \nodata &   & in Myr\\
$t_0$ & 500, 1000, 1500, 2000, 4000, 5000 & Age of the main stellar population in the galaxy in Myr \\
\hline
\multicolumn{3}{c}{Single-age stellar population (SSP): $\texttt{BC03}$ \citep{2003MNRAS.344.1000B} }\\
$Z$ & 0.05 & Metallicity \\
separation\_age & 10 & Age in Myr that separates the young and the old star  \\* &  & populations \\
\hline
\multicolumn{3}{c}{Nebular Emission \citep{2011MNRAS.415.2920I} }\\
$\log{U}$ & -2.0 & Ionisation Parameter \\
\hline
\multicolumn{3}{c}{Dust attenutaion: \citet{2000ApJ...533..682C} }\\
$\mathrm{E}(B-V)_{\mathrm{young}}$ & 0.1, 0.2, 0.3, 0.4, 0.5, 0.7, 0.9 &  Color excess of the stellar continuum light for the young population. \\
$\mathrm{E}(B-V)_{\mathrm{old factor}}$ & 0.44 & Reduction factor for the $\mathrm{E}(B-V)$ of the old population \\
  &  &  compared to the young one\\
$\delta$  & 0.0 &  Additional power-law index modifying the attenuation curve defined \\
  &  &  in Boquien et al. (2019)\\
\hline
\multicolumn{3}{c}{Clumpy two-phase torus model: $\texttt{SKIRTOR2016}$ \citep{2012MNRAS.420.2756S,2016MNRAS.458.2288S}} \\
$\tau$ & 5, 7 & The average edge-on optical depth at 9.7 micron \\
$\theta$ & 20, 40, 60 & The angle measured between the \\* & & equatorial plane and the edge of the torus \\  
$i$ & 0, 20, 50, 70 & The inclination of the line-of-sight  \\ 
$R_{out}/R_{in}$ & 20 & The ratio of the outer to the inner radii \\ 
fracAGN & 0.01, 0.1, 0.2, ... , 0.99 & The fraction of AGN IR luminosity in the total \\* & & IR luminosity \\  
\enddata
\end{deluxetable*}

\subsection{SED Fitting} \label{sec:xcigale}
Our samples were recorded in \citet{2016ApJS..224...24L} catalog, which provides $\rm{MASS\_BEST}$ (the stellar mass of the entire galaxy) and $\rm{SFR\_BEST}$ computed by the SED fitting. 
However, the SED fitting performed in this catalog do not include IR data of $Spitzer$ MIR channels, leading to large uncertainties in the SFR. 
Further, they did not take AGN contributions into consideration. 
Thus, to obtain results not only from the host galaxies but also the AGNs such as AGN luminosity and to update the catalog, we decided to perform our SED fitting.
We choose X-CIGALE \citep{2020MNRAS.491..740Y}, the X-ray version of CIGALE (Code Investigating GALaxy Emission, \citealt{2005MNRAS.360.1413B, 2009A&A...507.1793N, 2019A&A...622A.103B}), to perform SED fitting of the photometric data retrieved from COSMOS2015 catalog (see \S\ref{sec:sedata}). 

As the first step, a high-dimensional parameter grid of SED models consisting of all components that contribute to the emission was populated by X-CIGALE. 
Then, the goodness of the fit for each model was computed, and the best-fit SED model for each sample galaxy was identified via the reduced-$\chi^2$ statistic \citep{2020MNRAS.499.4325R}. 

The components used in the first step were required to be specified; we list these parameters and values in Table \ref{tab:2} to define the X-CIGALE grid of AGN hosts SEDs. 
For any other components not mentioned in the table, default settings provided by \citet{2020MNRAS.491..740Y} for AKARI and COSMOS AGN SEDs were used.

The star formation history (SFH) of the host was treated with a delayed SFH model \citep{2015A&A...576A..10C}: $SFR\propto \tau_{\rm main} \exp{(-t/\tau_{\rm main})}$.
Since the onset of the star forming activity, the increase in SFR is almost linear rather than a sudden burst; after a peak, the SFR decreases exponentially. 
There are two parameters controlling this model: the age from the onset of star formation ($t_0$), and the e-folding time ($\tau_\mathrm{main}$) that also determines the timing of the SFR peak. 
Through adjusting these two parameters, ongoing or recent starburst events as well as a quenched phase could be simulated. 

To model the emission from the stars, we adopted the \citet{2003MNRAS.344.1000B} stellar population synthesis library. The metallicity set by default was 0.02. 
However, considering that we were handling with AGN samples, we increased the value to 0.05 because AGNs would enrich their environments, resulting in higher chemical abundance \citep{2013MNRAS.431..793Z, 2015MNRAS.452L..59T}. 

Owing to the existence of the obscuring structure, emissions from the AGN will be re-emitted at longer wavelengths. 
To simulate this reddening effect, we adopted SKIRTOR, a clumpy two-phase torus model derived from a modern radiative-transfer method \citep{2012MNRAS.420.2756S,2016MNRAS.458.2288S}. 
This model depends on several parameters such as the average edge-on optical depth at 9.7 micron, the ratio of outer to inner radii, and the inclination. 
Because among our AGN host galaxies, \citet{2020ApJ...894...24K} suggested a significant fraction of optically unobscured Type 1 AGNs, we allowed these parameters to vary within wider ranges. 

\begin{figure*}[htbp!]
\begin{center}
\includegraphics[width=\textwidth]{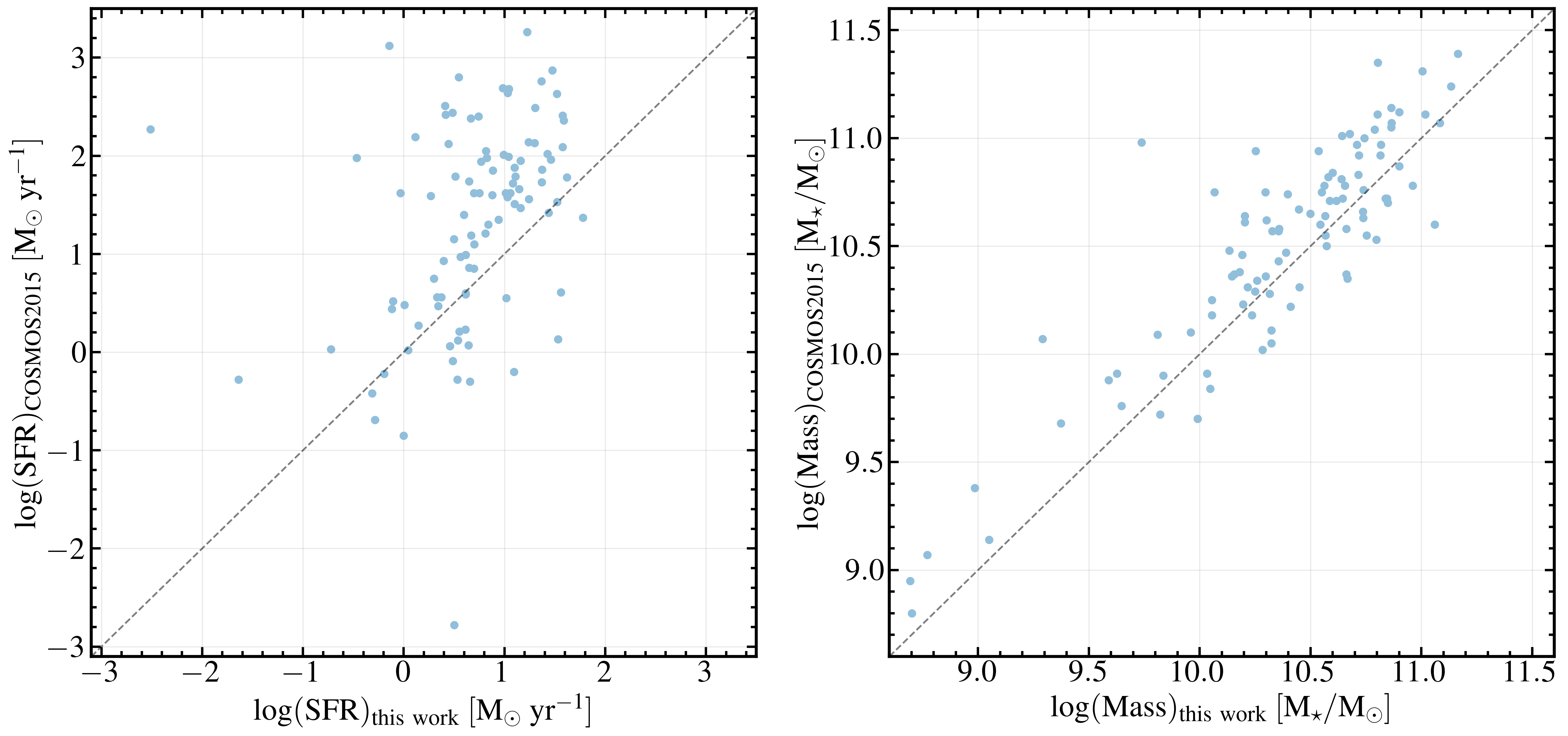}
\caption{\textcolor{black}{Star formation rates (SFRs, left panel) and stellar masses (right panel) in the COSMOS2015 catalog as a function of the values computed in this work.} \label{COSMOS2015}}
\end{center}
\end{figure*}

\begin{deluxetable*}{ccccc}[htbp!]
\tablecaption{Comparison of stellar masses and SFRs computed by X-CIAGLE and recorded in COSMOS2015.$^a$}
\tablehead{  & \multicolumn{2}{c}{log$_{10}$($SFR$/M$_\odot$ yr$^{-1}$)$_{\rm med}$} & \multicolumn{2}{c}{log$_{10}$($M_*$/M$_\odot$)$_{\rm med}$} \\
\colhead{Redshift} & \colhead{This work} & \colhead{COSMOS2015} & \colhead{This work} & \colhead{COSMOS2015}
}
\startdata
$z\leq1.0$ & \textcolor{black}{0.67}  & \textcolor{black}{1.51} & \textcolor{black}{10.33} & \textcolor{black}{10.57} \\
$1.0<z\leq1.5$ & \textcolor{black}{0.79} & \textcolor{black}{1.73} & \textcolor{black}{10.83} & \textcolor{black}{10.96} \\
$1.5<z$ & \textcolor{black}{0.95} & \textcolor{black}{1.95} & \textcolor{black}{10.85} & \textcolor{black}{10.75} \\
\enddata
\tablecomments{$^a$Three objects among our \textcolor{black}{$G-M_{20}+V$ classification samples} are not included in the comparison due to lack of SFRs and stellar masses in the COSMOS2015 catalog\label{tab:sedcom}.}
\end{deluxetable*}

\begin{figure}[htbp!]
\begin{center}
\includegraphics[width=0.47\textwidth]{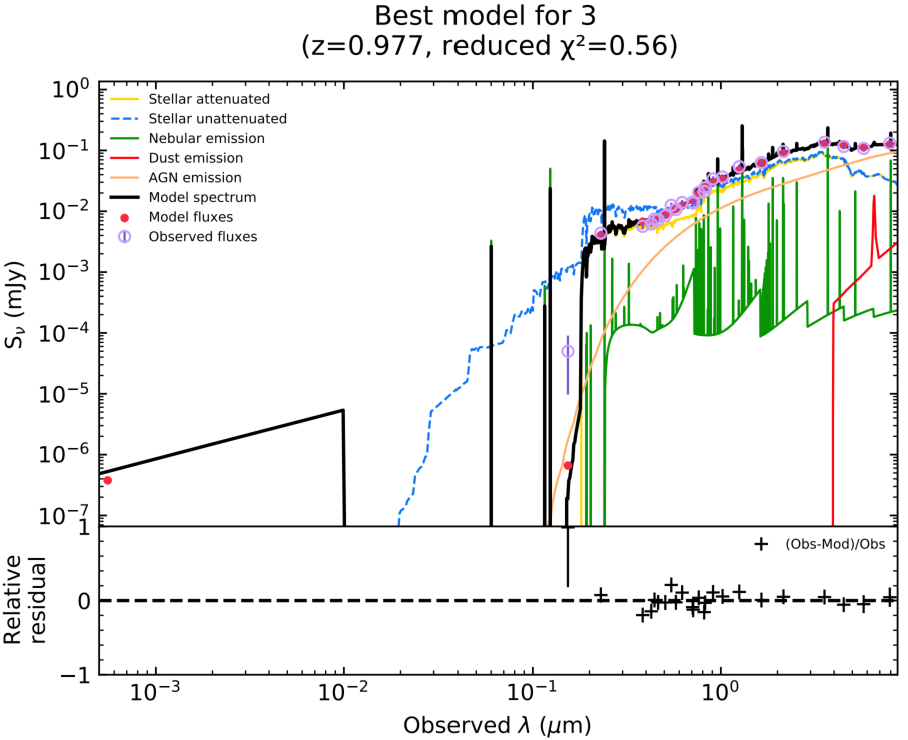}
\caption{An example of the best-fit models obtained from our SED fitting performed by X-CIGALE. This is the SED of the object ID003, a galaxy with visually merger-like morphology at $z=0.977$. The reduced-$\chi^2$ indicates the goodness of the fitting, and 0.56 ensures a high-quality and realistic fitting. The nebular emission contributes several emission lines in the $\texttt{Model spectrum}$ including the Ly$\alpha$ and H$\alpha$ line at $\lambda\sim0.24\mu$m and 1.3$\mu$m, respectibely, suggesting that the galaxy is undergoing star forming activities.\label{fig:ID3}}
\end{center}
\end{figure}

With the settings in Table \ref{tab:2}, we computed over 500 million models. 
In Table \ref{tab:sedcom}, we compare our median stellar masses and SFRs of \textcolor{black}{$G-M_{20}+V$ classification samples (see \S\ref{sec:diag}) with those in COSMOS2015 at three redshift bins, and Fig.~\ref{COSMOS2015} shows the corresponding values in the COSMOS2015 catalog as a function of our computed values.}
SFRs in COSMOS2015 show large scattering and are \textcolor{black}{2-dex higher than ours in the extreme case}, which can be a consequence of the absence of IR data and the ignorance of AGN components. 
The \textcolor{black}{medians} of the stellar masses are closer and most of the new estimates are 0.05--0.5-dex smaller than those in COSMOS2015. 
Fig.~\ref{fig:ID3} depicts the SED fitting results of the object ID003 at $z=0.977$ with visually merger-like morphology. 
The host galaxy is rich in emission lines, suggesting an ongoing star formation. 
The solid orange line indicates the emission from the AGN and AGN contributes a significant fraction to the observed fluxes, from which we may imply that this is a Type 1 AGN with a visible nucleus.
The computed SFR and inclination are $22.97\ \rm{M_\odot\ yr^{-1}}$ and $21.11^\circ\pm21.96^\circ$, respectively.

\begin{figure}[htbp!]
\begin{center}
\includegraphics[width=0.47\textwidth]{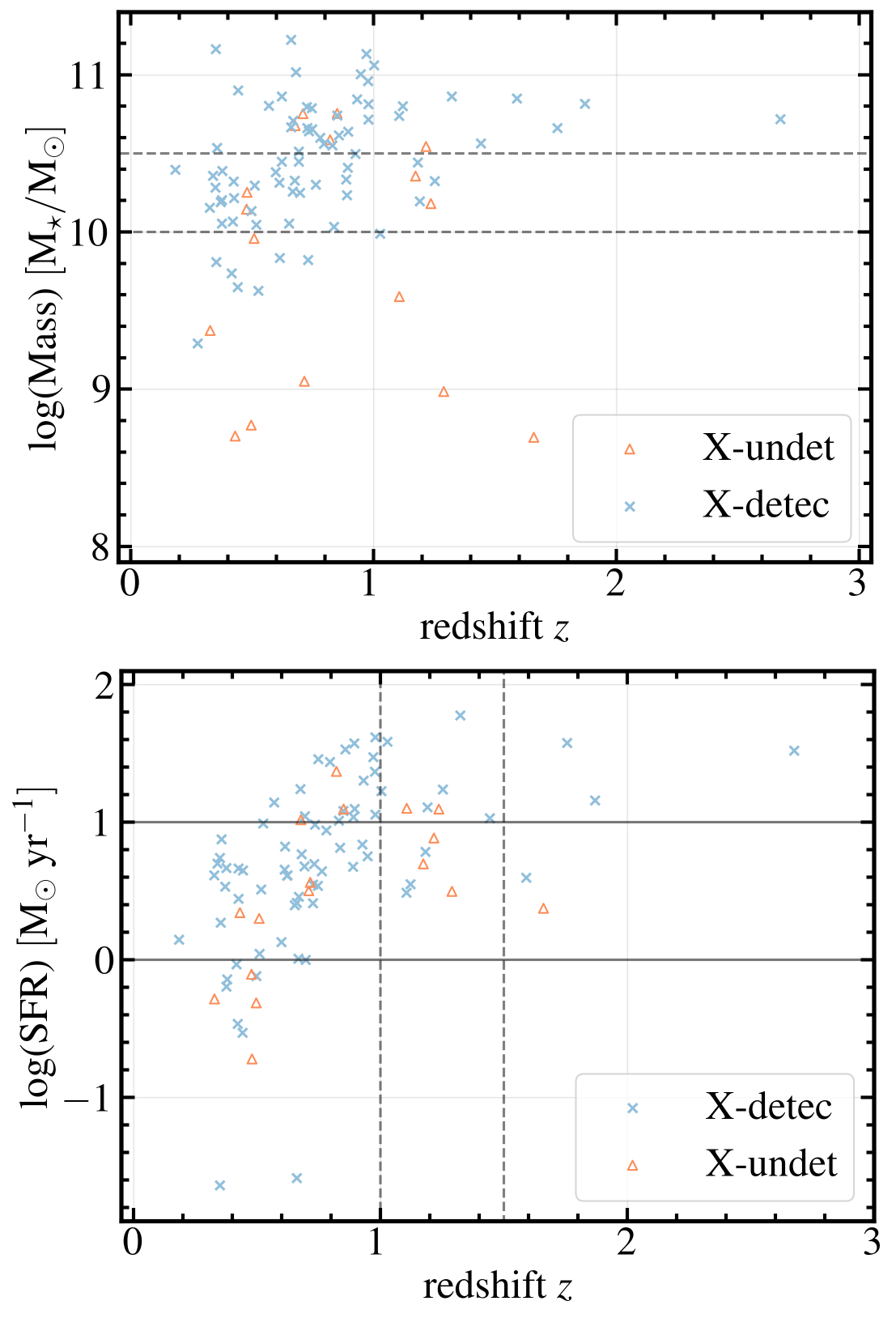}
\caption{Upper: Stellar mass of the entire galaxy computed by X-CIGALE as a function of the redshift. The dashed lines divide the sample into three bins along the stellar mass. Lower: SFR averaged over the recent 100 Myr computed by X-CIGALE as a function of the redshift. The solid and dashed lines divide the sample into three bins along the SFR and redshift, respectively. The blue crosses represent host galaxies with AGN detected by Chandra, while coral triangles are AGNs undetected in X-ray.\label{fig:z-mass,sfr}}
\end{center}
\end{figure}

\section{Results} \label{sec:results}
In this section, we will present the results of our morphological measurements with previous studies on AGN hosts and normal galaxies in the  literature. 
First, we describe our final sample to be discussed further. 

There were 491 objects in \citet{2020ApJ...894...24K}, and 485 of these were found in the HST-COSMOS database. 
By applying the selection criteria described in \S\ref{crit}, only 76 objects remained. 
There were other 26 objects that failed in passing the criteria due to invalid $\texttt{Gini\_segmentation}$ or outlying $r_e/sersic\_rhalf$, but their morphology could be confidently determined by the visual classification. 
Therefore, these 26 objects were also kept while the measurements of both parametric and non-parametric parameters were not discussed other than SED fitting results. 
Unfortunately, all AGNs at $z>3$ did not pass the criteria because the PSF-like component dominated the entire structure of their host galaxies.
Finally, there were only five valid objects at $1.5<z<3.0$.

For the 76 remaining objects found in the COSMOS2015 catalog, we computed the host galaxies' stellar masses and SFRs by using X-CIGALE as described in the previous section.
The redshift-mass and -SFR distributions are plotted in the upper and lower panel of Figures~\ref{fig:z-mass,sfr}, respectively. 
There are 72 ($\sim94.7\%$) host galaxies that lie within the star formation main sequence (see Fig.~\ref{fig:ssfr}), which can be considered as star forming galaxies (SFGs), with a median mass of $2.81\times10^{10} \mathrm{M_\odot}$.
\textcolor{black}{The central AGNs of these 76 objects have $L_{\rm {AGN,median}}\sim1.11\times10^{37}$ W, which is more than 1-dex fainter than the typical bolometric luminosity of samples in SDSS Quasar DR12 \citep{2017ApJS..228....9K}.
Therefore, we believe that these valid objects are good samples of low luminosity populations.}
In the following subsections of morphological measurements, we focus only on these 76 valid objects.

\begin{figure}
\begin{center}
\includegraphics[width=0.47\textwidth]{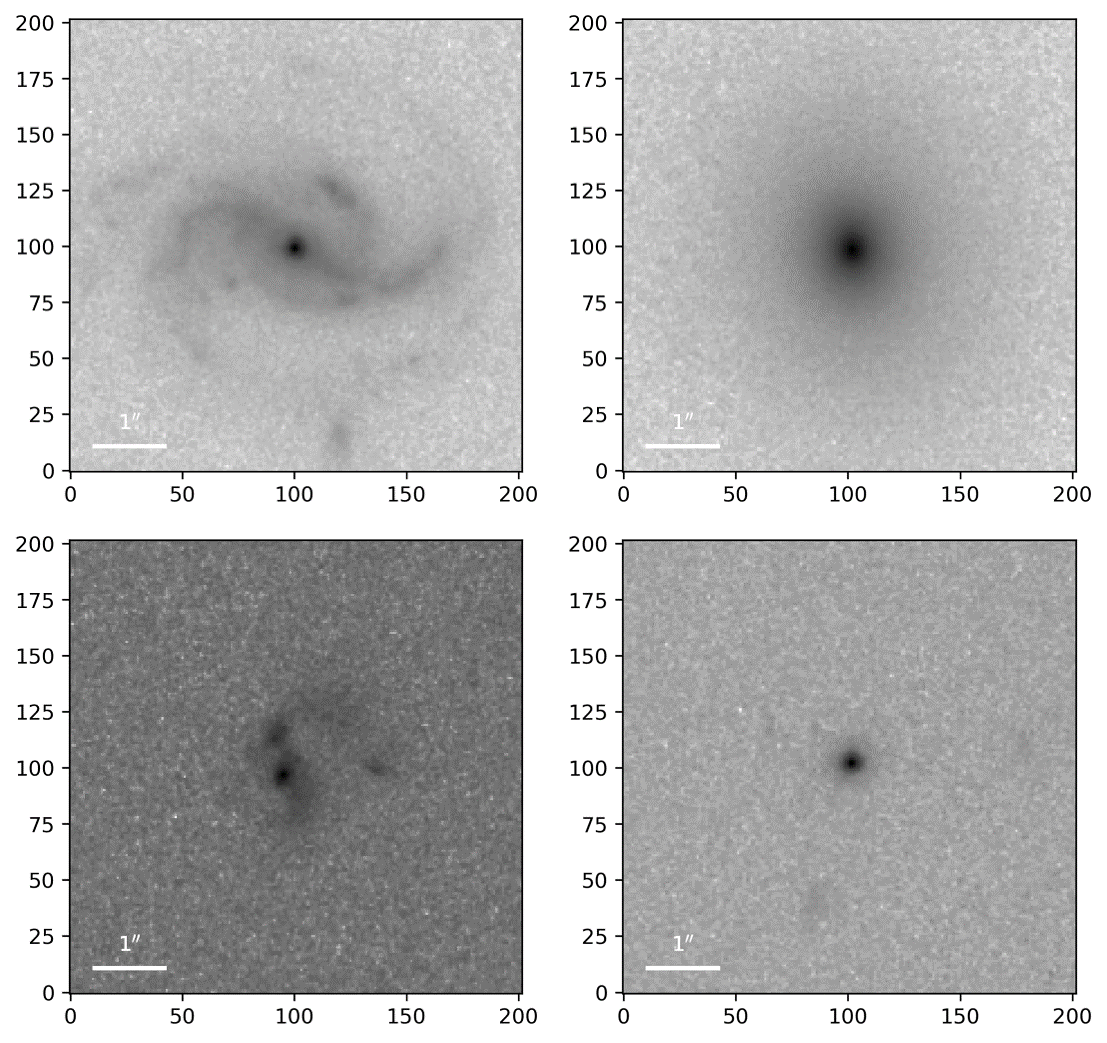}
\caption{Examples of visual classifications. From left to right and upper to lower, the galaxy is spiral, elliptical, irregular/merger, and point source, respectively. \label{fig:visualclass}}
\end{center}
\end{figure}

\begin{figure}[htp!]
\begin{center}
\includegraphics[width=0.47\textwidth]{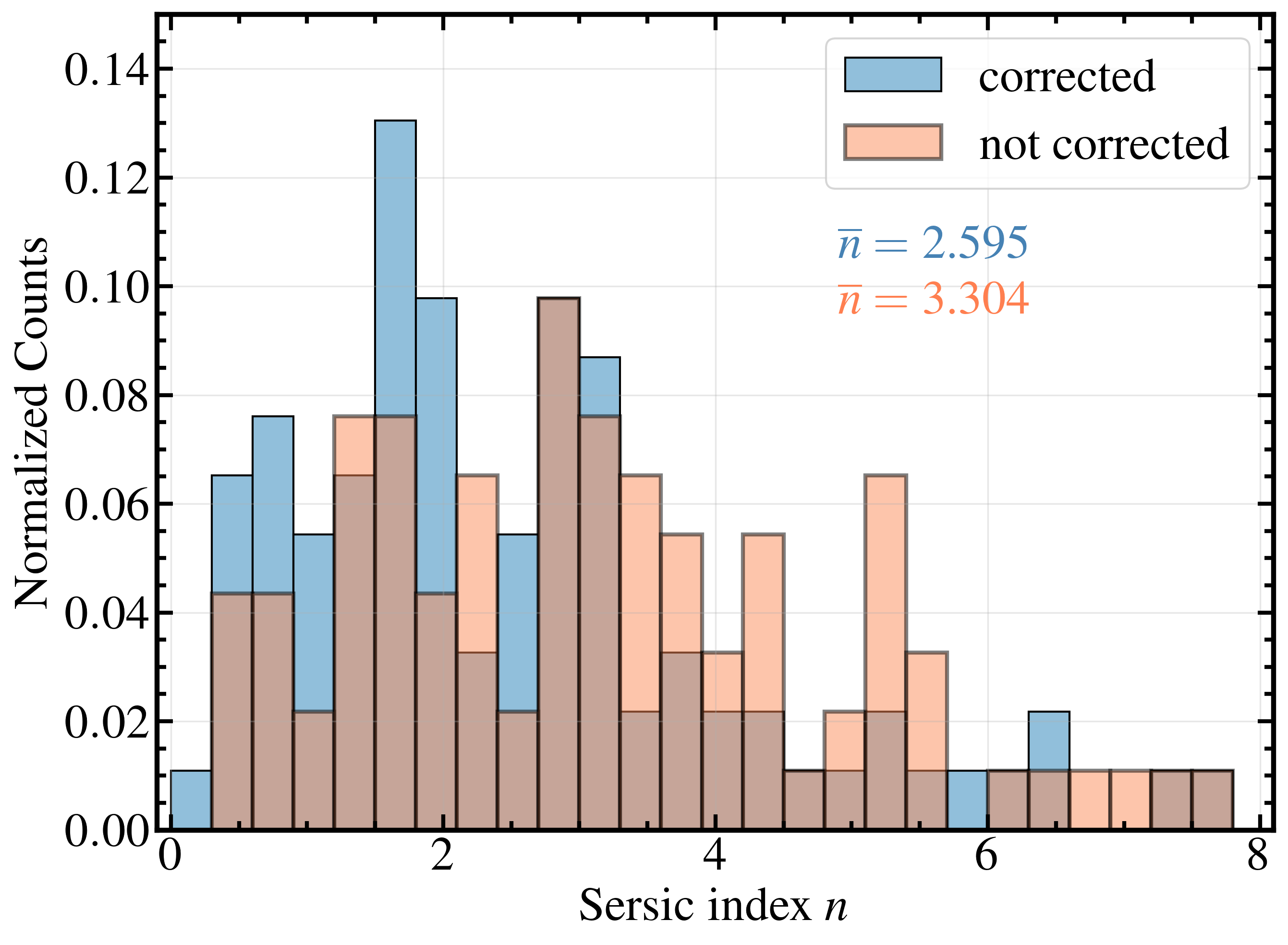}
\caption{The distribution of S\'ersic index before (coral) and after (steelblue) the correction. Without the correction, there is a clear bias toward higher values. This will result in an underestimation of the fraction of disk systems, thus leading to a conclusion of spheroid-dominance of the AGN hosts. After the correction, the valid sample shows no clear preferred morphological classification. \label{fig:calib}}
\end{center}
\end{figure} 

\begin{deluxetable}{ll}[htp!]
\tablecaption{A summary of visual classification for $76+26=102$ objects}
\tablehead{ \colhead{Classification$\ \ \ \ \ \ \ \ \ \ \ \ \ \ \ \ \ \ \ \ \ \ \ \ \ \ \  $} & \colhead{Number$^a$} }
\startdata
Spiral & $32$\\
Elliptical & $30$\\
Irregular/merger & $17$\\
Point-source & $23$\\
\enddata
\tablecomments{$^a$The pure visual inspection has uncertainties due to: spirals with patchy star forming regions may be classified as irregular/mergers and compact ellipticals lie on the boundary of ellipticals and point-source.  \label{tab:visual}}
\vspace{-1cm}
\end{deluxetable}

\subsection{Visual Classifications} \label{sec:visualclass}

Visual classification is the most original and explicit method to determine the morphology of a galaxy. 
However, owing to the difficulties in the classification of edge-on galaxies and those at high redshifts, where galaxies were more compact, we decided to follow the Galaxy Zoo 2 (GZ2, \citealt{2013MNRAS.435.2835W, 2011MNRAS.410..166L}) field guide to classify the total 102 objects into four: spirals, ellipticals, irregulars/merger, and point source.
Even after the rejection of point-like sources for which we could not obtain reliable morphological measurements as discussed in \S\ref{sec:correction}, we still found many small and compact sources among the valid objects, forcing us to classify them into the point source.
We showed examples of the four classes of galaxies in Fig.~\ref{fig:visualclass}. 
The result of the visual classification is listed in Table \ref{tab:visual}. 
Notably, there exist some objects that could not be confidently classified via a pure visual inspection.
These objects include point sources with spatial extensions of several pixels and are classified as ellipticals and galaxies with asymmetric arms but cannot be distinguished between spiral galaxies or late-stage mergers. 
Tiny arm-like substructures are also difficult to be distinguished between spiral arms or streams.

By visual investigation, many spiral galaxies showed some disturbed features such as asymmetric spiral arms and streams, which suggested that these galaxies were undergoing strong star formation, merger activities, or interactions. 
Owing to such difficulties, some visual classifications might be suspicious; we will discuss morphological classifications based on their S\'ersic index, Gini-$M_{20}$ and log(Gini)-log(Asymmetry) diagnostics in \S\ref{sec:diag}.

This visual method was also used as a criterion to decide the correction factor for the S\'ersic index and non-parametric parameters, as introduced in \S\ref{sec:correction}. 
Among the 76 valid objects, 26 had visible PSF components in their centers, corresponding to the modeled host galaxies that had their central regions not three times brighter than AGNs, and 50 did not have such visible PSF-like components. 

\begin{figure*}[htp!]
\begin{center}
\includegraphics[width=\textwidth]{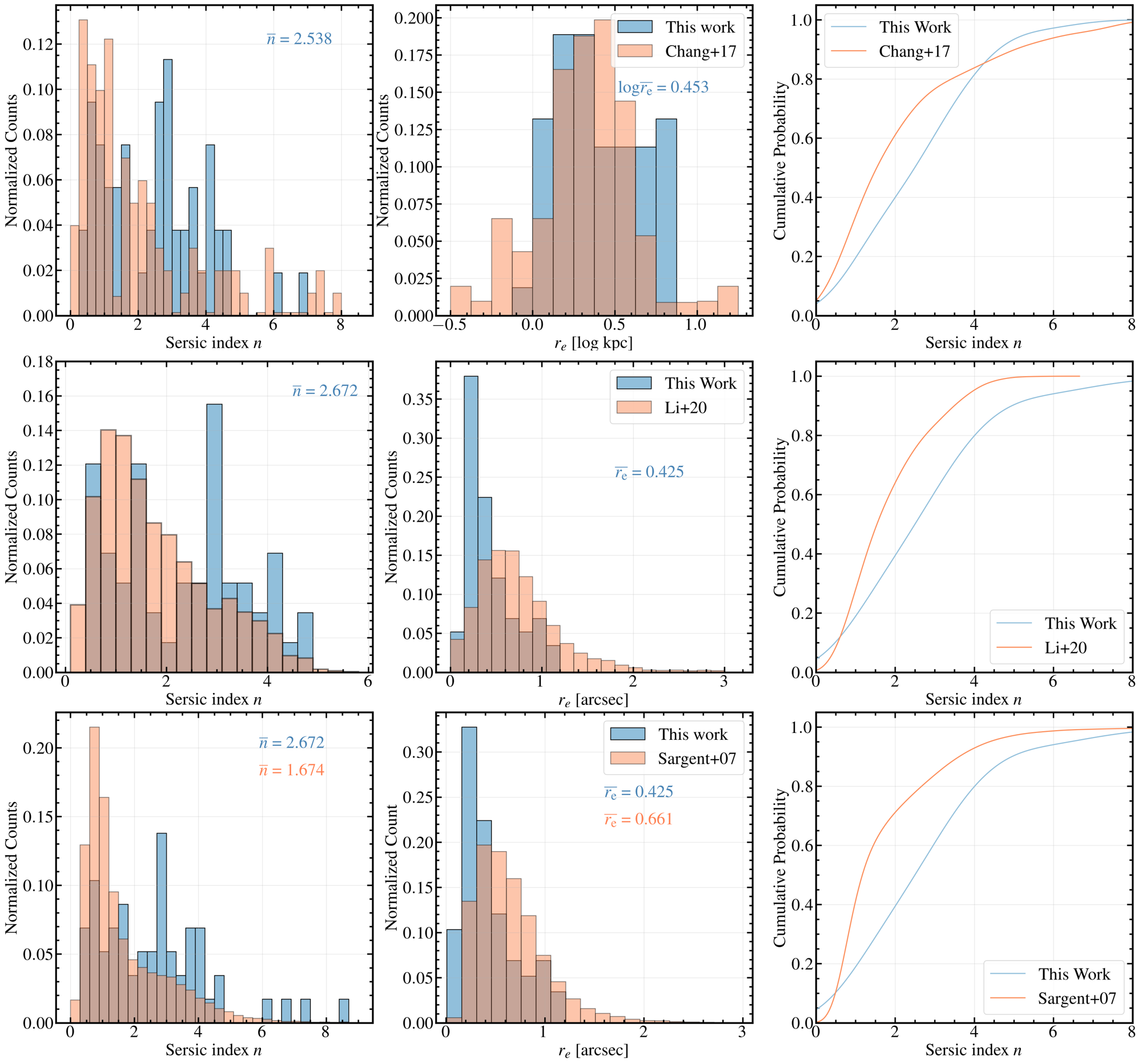}
\caption{Comparison of S\'ersic index and $r_{\rm e}$ in the literature. The redshift ranges of our samples match with the comparisons. Top: comparison with IR-selected AGNs at $0.5<z<1.5$ \citep{2017ApJS..233...19C}. Middle: comparison with quasars observed with Subaru HSC at $0.2<z<1.0$ \citep{2021arXiv210506568L}. Bottom: comparison with normal galaxies in the COSMOS field at up to $z\sim1$ \citep{2007ApJS..172..434S, 2007ApJS..172..406S}. $\Bar{n}$ and $\Bar{r_e}$ are average values of the S\'ersic index and the effective radius, respectively.
\label{fig:snr}}
\end{center}
\end{figure*}

\begin{figure*}[htp!]
\begin{center}
\includegraphics[width=\textwidth]{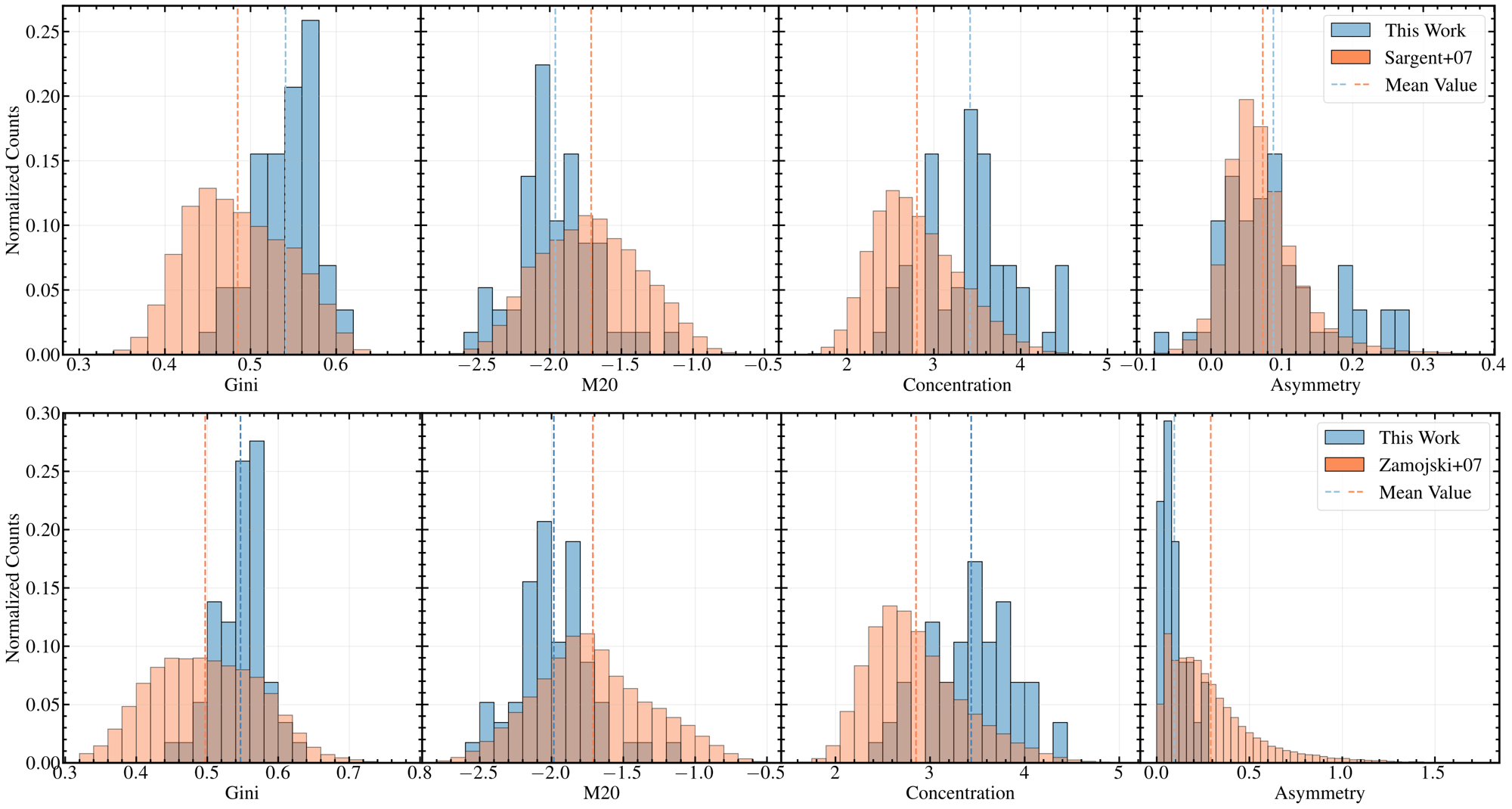}
\caption{Upper: $GMCA$ distributions compared with 16,358 normal galaxies ($\sim$12,000 disk galaxies) at $z<1$ \citep{2007ApJS..172..434S}; Lower: $GMCA$ distribution compared with 8,146 normal SFGs at $z\sim0.7$ \citep{2007ApJS..172..468Z}. Together with $GMC$, the host galaxies of our optical variability-selected AGNs had bright cores even compared with normal SFGs and were almost as symmetric as normal galaxies. \label{fig:gmcacosmos}}
\end{center}
\end{figure*}

\begin{figure*}[htbp!]
\begin{center}
\includegraphics[width=\textwidth]{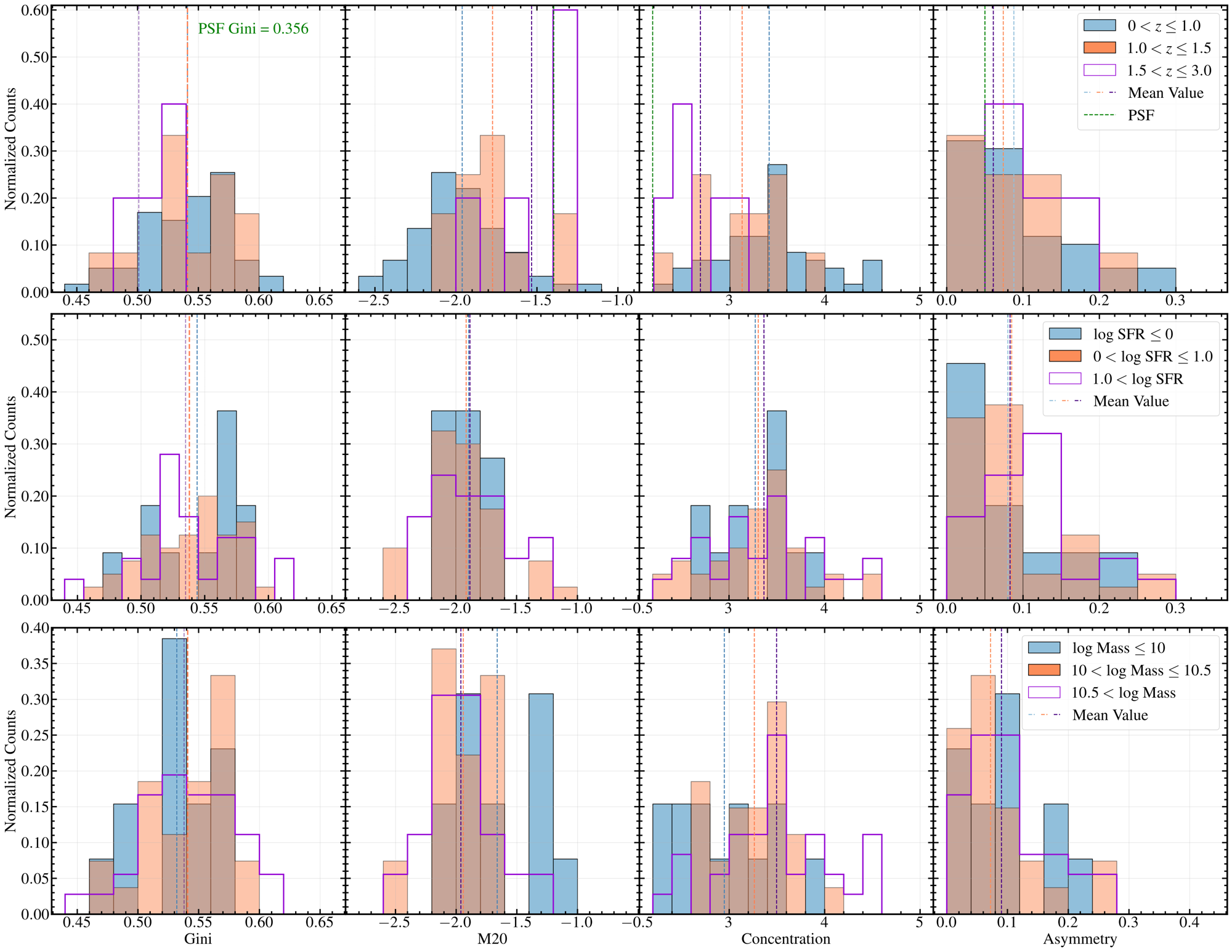}
\caption{Non-parametric parameters: Gini, $M_{20}$, Concentration, and Asymmetry of the variability-selected AGN host galaxies at three redshifts, total stellar masses, and SFRs bins. The Gini\_segmentation of the PSF only encloses the central core (smaller than 5*5 pixels), and the pixel counts within this small core are even (brightest $\sim0.6$, surrounding 8 pixels $0.3-0.5$), leading to both low Gini and Concentration. \label{fig:gmca}}
\end{center}
\end{figure*}

\subsection{S\'ersic Index and Effective Radius}
In this section we compare the results of 2D S\'ersic fitting with some other AGN host galaxy samples in the literature. 
First, we investigated the influence of the correction on distributions of the S\'ersic index.

The values of the S\'ersic index were corrected for bias due to the central bright AGN (\S\ref{sec:correction}). 
To clearly show this effect, we compared the values before and after the correction in Fig.~\ref{fig:calib}. 
The corrected histogram significantly shifted to smaller S\'esic index values than the histogram before correction.
Indeed, the S\'ersic index measurements without correction were biased toward higher $n$ values and the correction reduced the number fraction around $3<n<6$ significantly. 
Therefore, such correction to correct the bias is mandatory in morphological studies for AGN hosts to reject potential misleading dominant morphological classification.
Nevertheless, we discuss how this correction affects our conclusion about the dominant morphology. 
If we take $n=2$ as a standard to separate spheroid and disk systems, we find that 23 ($\sim30.3\%$) objects can be classified as disk systems before the correction and 36 ($\sim47.4\%$) after the correction. 
The conclusion changed from an apparent spheroid-dominance to an unapparent disk-dominance.

In the upper panel of Fig.~\ref{fig:snr}, our results are compared with the host galaxies of IR-selected AGNs within $0.5<z<1.5$, with $\mathrm{log} M_{\star}/M_\odot>10.5$ \citep{2017ApJS..233...19C}, whereas ours range from $\sim10^9\ \mathrm{M_\odot}$ to $\sim10^{11} \mathrm{M_\odot}$. 
Their AGNs were significantly obscured and the host galaxies could be considered as degradation, not contamination from AGNs. 
Considering $n=2$ \citep{2004ApJ...604L...9R,2011ApJ...743...96C} as the boundary that divided disk and spheroid systems, most host galaxies of IR-AGNs were disk systems in the top-left panel of Fig.~\ref{fig:snr}.
However, the S\'ersic index of the host galaxies of our optical variability-selected AGNs had a concentration at $n\sim3$ along with a higher cumulative probability ($\sim60\%$) at $n>2$. 
This led to a significant difference in the $n$ distribution of these two samples in a Kolmogorov–Smirnov test (KS test, P-value $\ll$ 0.01) . 
The distributions of $r_{\rm e}$ of these two samples were similar, but we found the absence of our objects in the most compact hosts ($r_{\rm e}\leq {\rm 0.42\ kpc}$) and extended ($r_{\rm e}>{\rm10\ kpc}$) cases as in the distributions of IR-AGN hosts.
This led to a less significant (P-value$\ \sim0.04$) but still a possible difference between the two samples. 
We explain this as that, our most compact hosts were dominated by the AGNs because of unobscuration and they failed to pass the criteria.
In addition, the average value of their host galaxies was about 3.56 kpc ($r_{\rm e}=10^{0.552}$ kpc, measured error considered), larger than ours ($r_{\rm e}=10^{0.438}=2.74$ kpc). 
However, considering the significant fraction of extremely extended IR-AGN host galaxies, we found no evidence of more compact sizes of hosts of the optical variability-selected AGNs.

We also compared the host galaxies of $\sim5000$ quasars from SDSS DR14 at $0<z<1.0$, $L_{bol}=10^{44.0-46.5}\mathrm{\ erg\ s^{-1}}$ and $9.5<\mathrm{log(M_\star/M_\odot)}<11.5$ \citep{2021arXiv210506568L} in the middle panel of Fig.~\ref{fig:snr}.
The morphology of these quasar host galaxies was studied using HSC five-band ($grizy$) optical imaging data.
Considering light contributions of quasars to the hosts, \citet{2021arXiv210506568L} also made corrections for the S\'ersic index by adding AGNs to the centers of galaxies with different magnitudes.
Apparently, the difference of $n$ between the two samples was highly significant (P-value $\ll$ 0.01).
However, in this comparison, our host galaxies showed a spheroid dominance while the majority of quasar host galaxies were still disk systems.
Besides, \citet{2021arXiv210506568L} have performed corrections for $n$ as well; for some quasar-dominated hosts fainter than HSC $i$-band$\sim23$, their intrinsic parameters could not be recovered. 
Such difficulty in the measurements of low luminosity hosts was also seen in our results before valid sample selection; over $\sim60\%$ host galaxies were fainter than $i_{\rm mag}\sim22.5$ and dominated by AGN components, and these objects had both $S=0$ and $\texttt{flag\_sersic}==1$. 
We removed all such objects and many of them even had a S\'ersic index of $n>10$.
Significantly different (P-value $<$ 0.01) distributions of $r_{\rm e}$ could also be seen that our sample was much more compact than the SDSS quasars.
We attributed this difference to the fact that \citealt{2021arXiv210506568L} used HSC imaging data. 
Because of the atmospheric seeing ($0.7''$), objects in HSC images were more extended, resulting in larger $r_{\rm e}$.

It is also interesting to mention the comparison with 16,538 normal galaxies selected with HST I-band magnitude ($I_{\mathrm{F814W}}<22.5$) from Zurich Structure \& Morphology catalog \citep{2007ApJS..172..434S, 2007ApJS..172..406S} in the bottom panel of Fig.~\ref{fig:snr}. 
Among these normal galaxies, $\sim$12,000 were disk systems. 
The distributions of $n$ significantly differed (P-value $\ll 0.01$), which was attributed to both the spheroid dominance in the AGN host galaxies and the disk systems that host AGNs had higher $n$ than normal disk systems.
For $r_{\rm e}$, our host galaxies were smaller in sizes (P-value $\ll 0.01$) and showed clearly separated concentrated regions.
Indicated by the average values, the size of our AGN hosts was roughly 70\% of that of normal galaxies. 

As suggested in \citet{2020ApJ...894...24K}, optical variability-selection is more efficient in selecting LLAGNs.
This was supported by the result of SED fitting that our AGNs had a median luminosity of $L_{bol}=1.11\times10^{44.0}\mathrm{\ erg\ s^{-1}}$ while SDSS quasars spanned $L_{bol}=10^{44.0-46.5}\mathrm{\ erg\ s^{-1}}$ with a median at $L_{bol}\sim10^{45.2}\mathrm{\ erg\ s^{-1}}$, which was larger that of ours by 1 dex \citep{2021arXiv210506568L}.
LLAGNs are likely to be hosted by SMBHs that have lower accretion rates, which are more likely to reside in the center of elliptical galaxies \citep{2014ARA&A..52..589H}.
This explains the different morphological dominance seen in the three comparisons as well as more compact sizes when they are compared with the sample of normal galaxies, 70\% of which are spiral galaxies \citep{2007ApJS..172..434S}.
The smaller sizes may also imply that our AGN-host SFGs are undergoing a process of dynamical compaction, probably arising from the gas inflow. 
Cosmological hydro-dynamical simulations of galaxy formation suggest that highly perturbed wet disks fed by cold streams may experience a dissipative contraction phase \citep{2014MNRAS.438.1870D}. 
In this scenario, the gas inflow, which is often associated with the disk instability, toward the central region of the galaxy triggers the initial contraction and acts as the energy-provider to maintain substantial high-level turbulence, leading to a massive core and enhanced SFRs, along with the triggering of the accretion onto the SMBH and induced AGN activities \citep{2015MNRAS.450.2327Z,2011ApJ...741L..33B}.

\subsection{Non-parametric Parameters} \label{sec:gmcaresult}

Non-parametric parameters provided more detailed investigations on the light distributions of our host galaxies. 
There were four parameters presented in the results: Gini ($G$), $M_{20}$, Concentration ($C$), and Asymmetry ($A$).

In the upper panel of Fig.~\ref{fig:gmcacosmos}, we plot the distributions of $GMCA$s of our AGN host galaxies and normal galaxies from \citet{2007ApJS..172..434S}. 
The correction described in \S\ref{sec:correction} was also applied to our measurements, although the AGN effects were small.
Apparently, normal galaxies had more even light distributions seen from their smaller $G$ values.
Combined with $M_{20}$ and $C$, it implied there were brighter central regions of the host galaxies of our AGN sample. 
Given that we corrected the bias due to the central AGN component in the measurements, the brighter central parts might serve as evidence of the central star formation induced by the AGN. 
As shown by Asymmetry, there was only a minute increase compared with normal galaxies, suggesting insignificant impacts of AGNs on the global structures of the host galaxies. 

In the bottom panel of Fig.~\ref{fig:gmcacosmos}, we compared the results with the COSMOS Zamojski Morphology catalog \citep{2007ApJS..172..468Z}. 
This catalog includes 8,146 SFGs selected based on HST ACS I-band with $I_\mathrm{F814W}\leq23$ at $0.2<z<1.0$ and $9.5<\mathrm{log}(M_\star/M_\odot)<11.6$. 
The distributions of $G$, $M_{20}$, and $C$ showed the same pattern as the normal galaxies of \citet{2007ApJS..172..434S}. 
Meanwhile, the distributions of $A$ differed. 
Their SFGs showd an extended tail to high $A$ values in the distributions, indicating an intense and inhomogeneous star formation in their disks.
Our AGN host galaxies were comparatively much more symmetric, which is also shown by the comparison in the upper panel. 
The significant fraction of spheroid systems implied by the S\'ersic index might be an explanation to the difference in the distributions of $A$.

We also plot the $GMCA$ measurements of our AGN samples within three redshifts, stellar masses, and SFR bins in Fig.~\ref{fig:gmca}. 
The average values of $G$ and $A$ at different redshift bins were close, whereas $M_{20}$ and $C$ showed a slight evolution pattern.
This could be explained as the result of the size evolution along the cosmic time, i.e., at higher redshifts, the galaxy became more compact, and spheroid systems showed a steeper decrease than disk systems \citep{2014ARA&A..52..291C}, as well as surface brightness dimming along the redshift \citep{2006ApJ...636..592L}. 
Therefore, with a smaller spatial extension, the contribution of the brightest pixel for high-$z$ objects did not influence the measurement as significantly as low-$z$ objects did.
The rest-frame UV light would be redshifted to the wavelengths at which we observed for high-$n$ objects. 
Therefore, we expect more UV light for higher-$z$ galaxies. 
Because the star forming regions are the main contributor in UV, more disturbed and asymmetric structures can be expected as in \citet{2007ApJS..172..468Z}. 
However, such asymmetry evolution along the redshift was not clearly seen in $A$.

\begin{figure}[htbp]
\begin{center}
\includegraphics[width=0.47\textwidth]{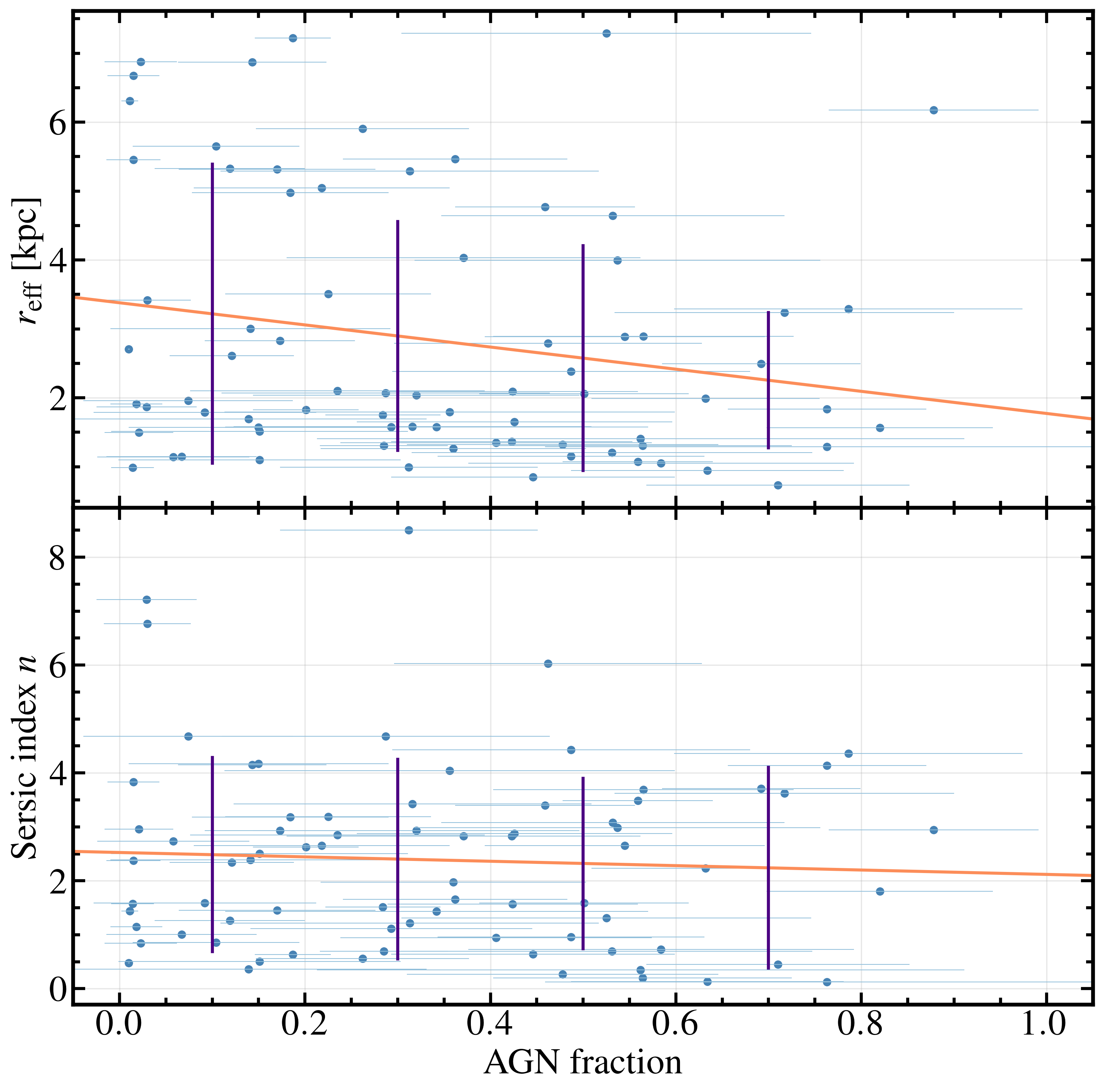}
\caption{The effective radius, $r_{\rm e}$, and S\'ersic index, $n$, as functions of the AGN fraction calculated from IR luminoisty. The coral lines are linear fitting and the indigo lines indicate standard deviations within different bins of the AGN fraction. \label{fig:fracAGN}}
\end{center}
\end{figure}

Comparisons of different stellar mass and SFR bins are shown in the middle and lower panels of Fig.~\ref{fig:gmca}, respectively. 
We could barely see any differences between $G$, $M_{20}$, and $C$ for both comparisons, indicating that the stellar masses and star formation had no clear dependencies on the global morphology. 
Besides, as seen in the distribution of $A$ for different SFR bins, although there was an increase in $A$ toward higher SFRs, the evolution was quite insignificant compared with normal SFGs, whereas these hosts of optical-variability selected AGNs were also SFGs. 
Such results again required a morphological classification to explain and suggested that the star forming activities in these hosts might more likely to be related to the galaxies' central regions.

\subsection{AGN Fraction Contribution}

Through the SED fitting, we also computed the AGN fraction defined as the AGN luminosity fraction in the total (AGN$+$galaxy) IR luminosity \citep{2019A&A...622A.103B,2020MNRAS.491..740Y}. 
In this section, we investigate the relations of the AGN fraction with the S\'ersic index, $n$, and the effective radius, $r_{\rm e}$, to discover whether these parameters correlate with the AGN fraction. 

Fig.~\ref{fig:fracAGN} shows the relations between the AGN fraction and the S\'ersic index, $n$ as well as the effective radius, $r_{\rm e}$.
The effective radius $r_{\rm e}$ and S\'ersic index showed negative correlations with the AGN fraction. 
Their linear-fittings with the AGN fraction are given by: $r=-(1.61\pm0.91)f+(3.38\pm0.37)$ and $n=-(0.41\pm0.85)f+(2.52\pm0.35)$, where $r$ is the effective radius, $n$ is the S\'ersic index, and $f$ is the AGN fraction. 
According to these equations, a 20\% AGN contribution led to a 3.3\% decrease of the S\'ersic index and 9.5\% decrease in the $r_{\rm e}$ compared with those at a zero AGN fraction, respectively. 
A 50\% AGN contributions corresponded to an increased S\'ersic index by 8.1\% and a decreased $r_{\rm e}$ by 23.8\% compared with the case without any AGN effect.

To measure the significance of these correlations, we calculated Pearson's correlation coefficients. 
For $r_{\rm e}$, it had a P-value of $0.08>0.05$ and a correlation coefficient of $r=-0.2$, and for S\'ersic index, the P-value and correlation coefficient were $0.63\gg0.05$ and $r=-0.06$, respectively.
It suggested that $r_{\rm e}$ is weakly correlated with the AGN fraction and the correlation was not statistically significant.
This weak correlation between $r_{\rm e}$ and the AGN fraction agreed with the process of dynamical contraction that arising from the central gas inflow, 
Meanwhile, the S\'ersic index, in principle, did not correlate with the AGN fraction at all, which might be because we corrected the influences of AGN components and removed extremely compact objects.
We calculated the P-values of the AGN fraction between non-parametric parameters. 
However, all of them had a P-value $\gg0.05$, indicating no arguable correlations between the AGN fraction and non-parametric parameters.

\begin{figure*}[htbp!]
\begin{center}
\includegraphics[width=\textwidth]{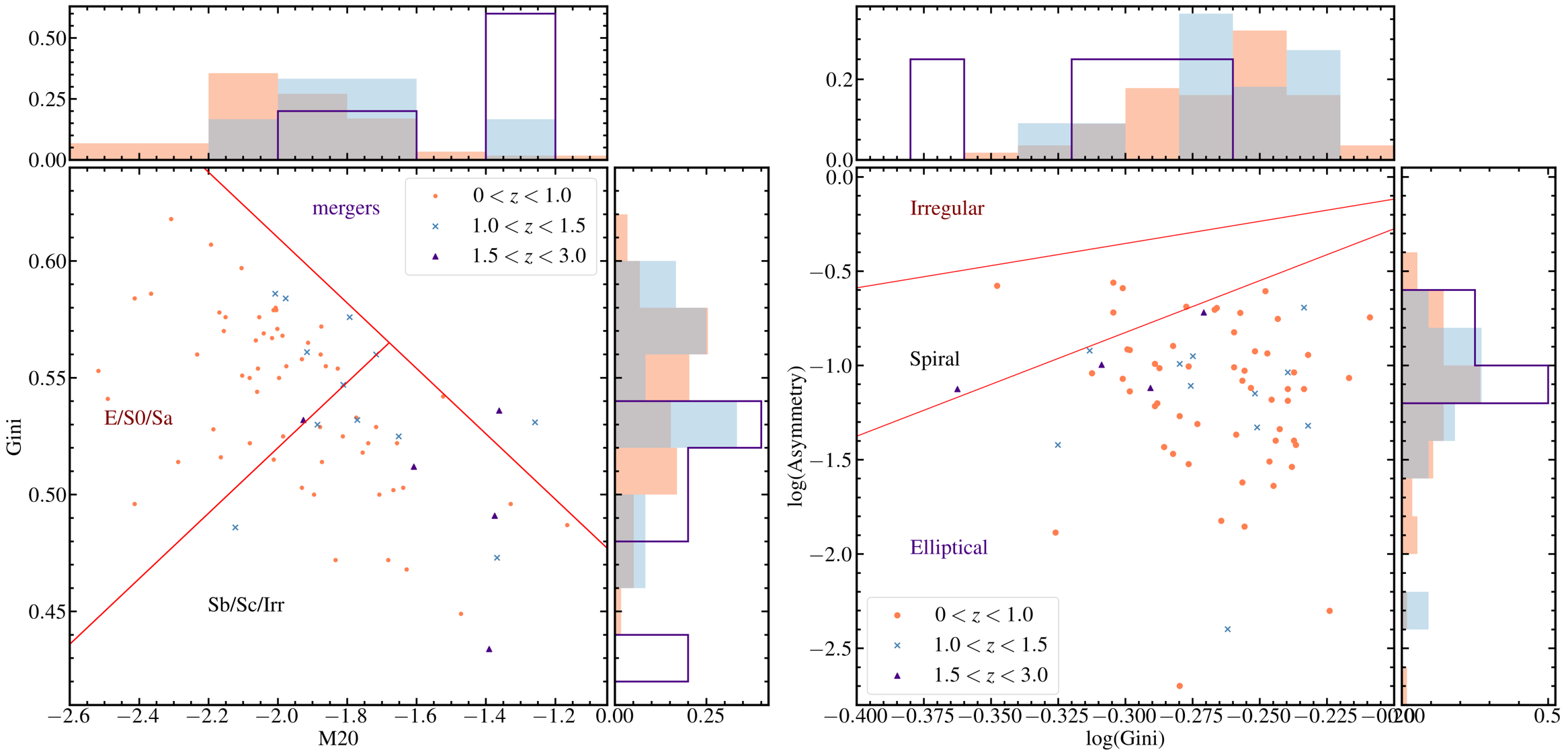}
\caption{Left panel: Gini-$M_{20}$ diagnostics that puts galaxies into three classification regions: merger, E/S0/Sa, and Sb/Sc/Irr. The division lines are obtained from \citet{2008ApJ...672..177L}. Right panel: log(Gini)-log(Asymmetry) diagnostics that roughly classify the morphology with three basic types: irregular, spiral, and elliptical. We adapted the formula in \citet{2007ApJS..172..284C} to divide the regions. Gray histograms in the distribution simply show the overlapped objects. \label{fig:diagnostics}}
\end{center}
\end{figure*}

\section{Discussion} \label{sec:discus}

\subsection{Non-parametric Parameter Diagnostics}\label{sec:diag}

Other than visual classification, a method to determine which class the galaxy belongs to is the diagnostics based on non-parametric parameters, including Gini-$M_{20}$ ($G-M_{20}$) and log(Gini)-log(Asymmetry) (log($G$)-log($A$))  diagnostics. 

In the left panel of Fig.~\ref{fig:diagnostics}, we plot the $G-M_{20}$ diagnostics that divides the region into three parts: E/S0/Sa, Sb/Sc/Irr, and Mergers. The division lines are obtained from \citet{2008ApJ...672..177L} with the following definitions for Extended Groth Strip (EGS) galaxies at $0.2<z<1.2$: $\mathrm{Mergers}: G > -0.14 M_{20}+0.33$, $\mathrm{E/S0/Sa}: G\leq -0.14  M_{20}+0.33~ \& \ G > 0.14M_{20}+0.80$, and $\mathrm{Sb/Sc/Irr}: G \leq -0.14M_{20}+0.33~ \& \ G \leq 0.14M_{20}+0.80$. 
We show the fraction of different identifications at different redshifts in Table \ref{tab:gm+v}.
Clearly, most AGN host galaxies at low redshifts resided in the E/S0/Sa, whereas at higher redshifts, the probability of them being found in the Sb/Sc/Irr region increased. 
Besides, the fraction of mergers was small within the full redshift range, suggesting that mergers were not the major mechanism triggering AGN activities, consistent with the findings of \citet{2017ApJS..233...19C}. 
However, as the redshift increased, the merger fractions increased as well, which may might cause a sudden strong gas inflow toward the galaxy center and eventually triggers the AGN activity. 
In our visual classification, many spiral galaxies showed disturbed features, which might support this implication.

The log($G$)-log($A$) diagnostics is plotted in the right panel of Fig.~\ref{fig:diagnostics}. 
The total sample had a size of 71 objects since five objects had negative $A$ values due to high background noises. 
The division lines were defined by \citet{2007ApJS..172..284C}, who studied the morphology of the galaxies in the COSMOS field at $0<z<1.2$: the division line between the irregular and spiral is: $\log_{10}A= 2.353\cdot\log_{10}G+0.353$, and the division line between the spiral and elliptical is: $\log_{10}A=5.50\cdot\log_{10}G+0.825$. 
In the entire redshift range, 65 ($\sim91.5\%$) were classified as elliptical galaxies, and only six objects ($\sim8.5\%$) were spiral galaxies. Further, there were no irregular or merging galaxies. 
The number of spiral galaxies in this diagnostics is even smaller than that of visually confirmed spirals. 
We infer that such distributions could be attributed to the low $A$ and larger \textcolor{black}{$G$} at the same time. 
The $A$ of these AGN hosts had only slight increase compared with normal galaxies, whereas they had much more uneven light distributions.

Albeit in $G-M_{20}$ diagnostics, 46 objects ($\sim60.5\%$) were E/S0/Sa, since it included S0 and Sa, we expected a smaller fraction of ellipticals.
Further, the 2D S\'ersic fitting depends on the underlying mathematical form, which means it cannot be used as an indicator of merging/irregular galaxies.
In addition, when we handled these systems, errors might occur and reduce the valid sample size.
Therefore, we combined the $G-M_{20}$ diagnostics and visual classifications ($G-M_{20}+V$) to reach the final morphology result for valid objects, which included objects that were successfully measured and objects with visually recognizable morphological structures but failed to pass the selection criteria. 
In $G-M_{20}$ diagnostics, 17 of 22 visually classified ellipticals were classified as E/S0/Sa, whereas only 10 of 20 visually classified spirals were classified as Sb/Sc/Irr.
Among the other 10 visual spiral galaxies, six had two visible spiral arms, by which they were classified as Sa. 
We took the results of $G-M_{20}$ diagnostics as the priority and visually reinvestigated the objects with inconsistent classifications between the two methods to avoid misclassifications of $G-M_{20}$ diagnostics because of unenclosed faint substructures.
The visually-confirmed S0/Sa galaxies were all classified as disk systems, whereas the left ones were classified as spheroid systems.
The final classifications of $G-M_{20}+V$ is listed in Table \ref{tab:gm+v}.

As seen in the $G-M_{20}$ diagnostics, the dominant system is spheroid system at $z<1.5$.
Although at $z>1.5$, disk galaxies had a higher fraction, because of the limited sample size, this dominance was unarguable.
At $z\leq1$, the ellipticals occupied the absolute majority and the number of elliticapls was even over half of all ellipticals within the entire redshift range.
One possible explanation for this result could be attributed to less UV light, which primarily arose from star forming activities, received by the $F814W$ filter at lower redshifts. 
In addition, spheroid systems experienced a size evolution that dropped more steeply along the cosmic time than disk systems, resulting in a smaller fraction of detectable ellipticals with a higher redshift.
This was clearly shown in the valid sample selection that $\sim70\%$ host galaxies were excluded because of PSF-dominance, caused by small $r_{\rm e}$, large $n$, and bright AGN components.  

\begin{deluxetable}{cccc}[htp!]
\tablecaption{Summary of the classification of AGN hosts.\label{tab:gm+v}}
\tablehead{ \colhead{redshift} & \colhead{E/S0/Sa} & \colhead{Sb/Sc/Irr} & \colhead{Mergers}}
\startdata
\multicolumn{4}{c}{$G-M_{20}$ diagnostics$^a$}\\
$0<z\leq1.0\ (59)$ & 66.1\% & 20.0 \% & 0\% \\
$1.0<z\leq1.5\ (12)$ & 50.0\% &  41.7\% & 8.3\% \\
$1.5<z\leq3.0\ (5)$ & 20.0\% & 60.0\% & 20.0\% \\
$0<z<3.0\ (76)$ & 60.5\% & 36.8\% & 2.6\% \\
\hline
\hline
redshift & spheroid & disk & merger \\
\hline
\multicolumn{4}{c}{$G-M_{20}$ diagnostics + Visual classification$^b$}\\
$0<z<3.0\ (102)$ &  46.1\% & 44.1\% & 9.8\% \\
\enddata
\tablecomments{$^a$Numbers of objects within each redshift bin are shown in parentheses; $^b$Twenty-six objects with visually confident morphology that did not pass the selection criteria are included in $G-M_{20}+V$ classification.}
\vspace{-1cm}
\end{deluxetable}

\begin{figure}[htbp!]
\begin{center}
\includegraphics[width=0.47\textwidth]{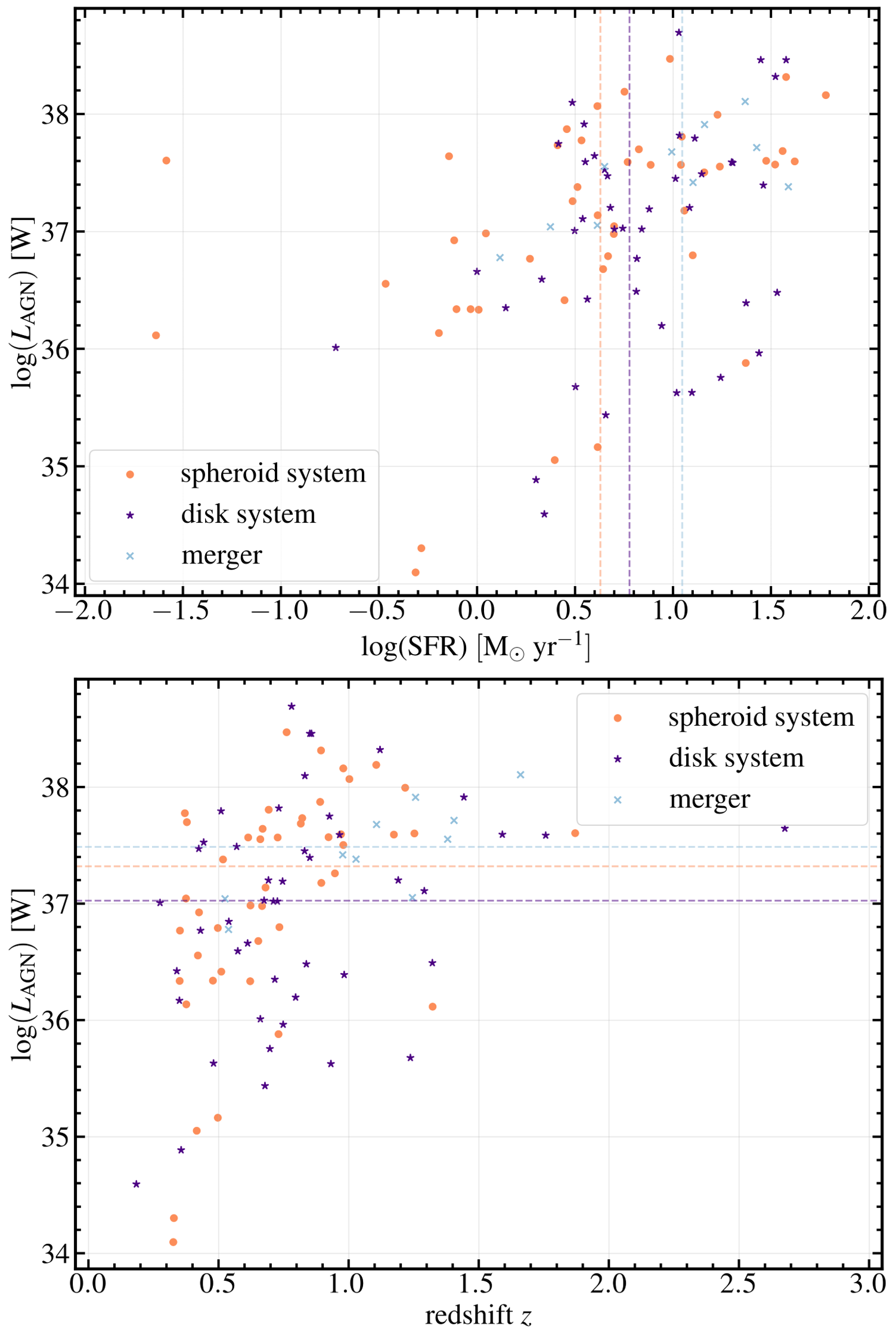}
\caption{Upper: AGN luminosity as a function of the SFR. Lower: AGN luminosity as a function of redshift. The meaning of the symbols are indicated in the panel. For both panels the classifications are based on the combination of the $G-M_{20}$ diagnostics and visual classification. The dashed lines in the upper and lower panels indicate the median values of log(SFR) and log($L_{\rm AGN}$), respectively. \label{fig:lagn}}
\end{center}
\end{figure}

\begin{figure}[htp!]
\begin{center}
\includegraphics[width=0.47\textwidth]{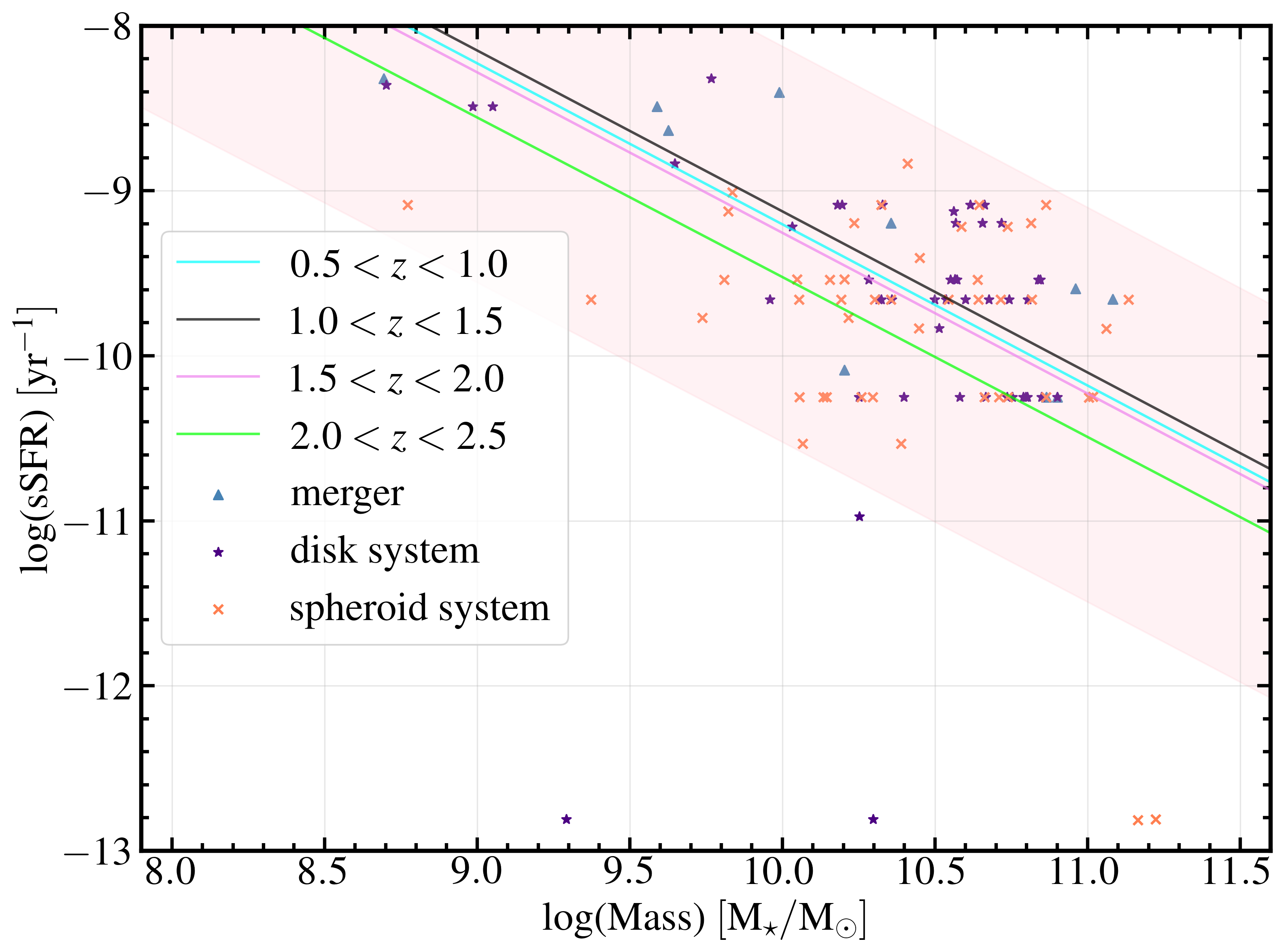}
\caption{\textcolor{black}{Specific star formation rate (sSFR)} as a function of galaxy mass. Different types of galaxies represented by different shapes are based on $G-M_{20}$ diagnostics + visual classification. The different lines show the SFR main-sequences at different redshifts adapted from \citet{2017ApJS..233...19C}\label{fig:ssfr}}
\end{center}
\end{figure}

\subsection{Implications to Galaxy-AGN Co-evolution}

As shown in the upper panel of Fig.~\ref{fig:lagn}, the majority of the hosts of our variability-selected AGNs were undergoing star forming activities. 
To investigate if they had stronger SFR than normal SFGs that lay around the star formation main sequence at the corresponding epochs, we plot the sSFR against the stellar mass in Fig.~\ref{fig:ssfr}. 
These star forming AGN hosts lie within the main sequence if $\pm1.0$ dex is considered. 
This suggested that there was no intensively enhanced SFR on the spiral galaxies that were intrinsically SFGs. 
But interestingly, for ellipticals, which were representative of quiescent galaxies with low SFRs, we had found that they were around the star formation main sequence, which suggested that their star forming activities might be triggered by AGNs. 
Such star forming elliptical galaxies hosting AGNs might answer a strange pattern shown in the Asymmetry comparison in the lower panel of Fig.~\ref{fig:gmcacosmos}, i.e., star forming AGN host galaxies showed more symmetric structures than normal SFGs. 
The Asymmetry within different SFR bins in the lower panel of Fig.~\ref{fig:gmca} is a complement to this pattern; higher SFRs only have negligible impacts on the global structures of the host galaxies. 
However, there was an almost equal fraction of disk systems, which should be more asymmetric than ellipticals and show a long tail in Asymmetry but contribute slightly to the Asymmetry evolution. 
Similarly, \citet{2009ApJ...691.1005R} studied $\sim25,000$ nearby galaxies ($z<0.06$) and compared the lopsidedness of non-AGN and AGN galaxy pairs matched in redshift, mass, mass density, and stellar age but found no significant difference in the lopsidedness of matched galaxy pairs. 
We interpret the above results the evidence that AGN feedback affects the SFR primarily in the central regions of the host galaxies and its influence on the entire system is less significant.
This interpretation is also supported by \citet{2013ApJ...765L..33L}, who studied 28,000 Type 2 Seyfert galaxies at $z\lesssim0.3$ and found positively correlated AGN luminosity and centrally concentrated star formation.

Other than the simple classification of $G-M_{20}$ diagnostics, our visual classification revealed that a relatively larger fraction of spiral galaxies had disturbed features. 
This indicated the existence of possible interactions or merger activities and conflicted with \citet{2011ApJ...726...57C}, who studied the hosts of X-ray-selected Type 1 AGNs at $z\sim0.3-1.0$ and found that 85\% of their objects showed normal undisturbed morphological patterns. 
Owing to the limitation of observations, we cannot give certain words that these features were attributed to disk instabilities or that they were late-type mergers. 
Except for the disturbed galaxies, we also found 10 mergers or irregular galaxies in the imaging data.
The basic conclusion for the small merger rate at $0<z<3.0$ agreed with many previous studies arguing that mergers were not the primary triggering mechanism of AGN activities \citep{2008ApJ...688...67E, 2012ApJ...757...81B}. 
However, in the lower panel of Fig.~\ref{fig:lagn}, we did find that mergers were more likely to be found at higher redshifts, which agreed with the predictions of cosmological hydrodynamic simulations \citep{2016MNRAS.462..190R}.
We also checked the AGN fractions, AGN luminosities, SFRs, and stellar masses of host galaxies, and we have found that, with higher SFR and AGN luminosity, the galaxy had a larger possibility to be a merger-like object, as shown in the bottom panel of Fig.~\ref{fig:lagn}. 
Such finding agreed with the X-ray-selected and HST/WFC3 imaged heavily obscured AGNs at $z\sim1$ \citep{2015ApJ...814..104K}. 
This can be understood as an aspect of the evolution of the massive galaxy, i.e., the most massive systems grow mainly through hierarchical merging activities. 
This is more ubiquitous at younger cosmic times relative to the local universe. 
Then, the merging galaxies induce a sudden gas inflow into their central regions that feed strong AGN activities compared with steady gas inflow. 
Intensive SFRs and luminous AGNs are expected as the products of these major mergers.

\section{Conclusions} \label{sec:conc}
In this study, we studied the morphology of host galaxies with optical variability-selected AGNs at $0<z<3$ in the COSMOS field. 
The host morphology was evaluated using parametric ($n$) and non-parametric ($G$, $M_{20}$, $C$, $A$) morphological parameters and investigated with SFRs using HST imaging ($\sim0.03''$) and SED fitting. 
Our main conclusions are as follows.

\begin{enumerate}
\item The sizes of host galaxies of the optical variability-selected AGNs up to $z\sim1$ are more compact than normal galaxies at the same redshift and stellar mass range by 35.7\%.
\item The host galaxies of these optical variability-selected AGNs have no clear morphological preference, seen in the number fractions of the disk ($\sim44.1\%$) and spheroid ($\sim46.1\%$) systems.
\item Almost all AGNs ($\sim94.6\%$) reside in SFGs ($\mathrm{log(SFR)_{med}\sim0.7\ M_\odot\ yr^{-1}}$) that have a very similar Asymmetry distribution compared with that of normal galaxies, and much more symmetric structures compared with normal SFGs. This can be explained by the fraction of elliptical galaxies ($44.9\%$), suggesting that the AGN feedback enhances the star formation of spheroid systems and the star forming activities influenced by AGN feedback only vary within a small central scale rather than the entire system.
\item The fraction of major mergers in the variability-selected AGNs is as small as $\sim9.8\%$, which suggests that major mergers are not the main triggering mechanism of AGN activities; however, the merger rate increases with the increase in the redshift, AGN luminosity, and SFR values.
\end{enumerate}

\acknowledgements
We thank Yu-yen Chang, Mark Sargent, Michel Zamojski, and Junyao Li for providing their data in literature and the referee for the useful comments.
This work would have not be possible without Vicente Rodriguez-Gomez’s technical support.
This work used the NASA/IPAC Infrared Science Archive (IRSA) at the NASA Infrared Processing and Analysis Center (IPAC), located on the campus of the California Institute of Technology (Caltech).
We gratefully acknowledge the contributions of the entire COSMOS collaboration. 
The COSMOS team in France acknowledges support from the Centre National d’\'Etudes Spatiales.\\
$Software$: statmorph \citep{2019MNRAS.483.4140R}; Photutils \citep{larry_bradley_2020_4044744}; Tiny Tim \citep{2011SPIE.8127E..0JK}; Astropy \citep{astropy:2013, astropy:2018}; X-CIGALE \citep{2019A&A...622A.103B, 2020MNRAS.491..740Y}.

\bibliography{paper2}
\bibliographystyle{aasjournal}

\appendix

\section{Correction formula for the S\'ersic index}\label{apd}

\begin{figure}[htbp!]
\begin{center}
\includegraphics[scale=0.28,angle=0]{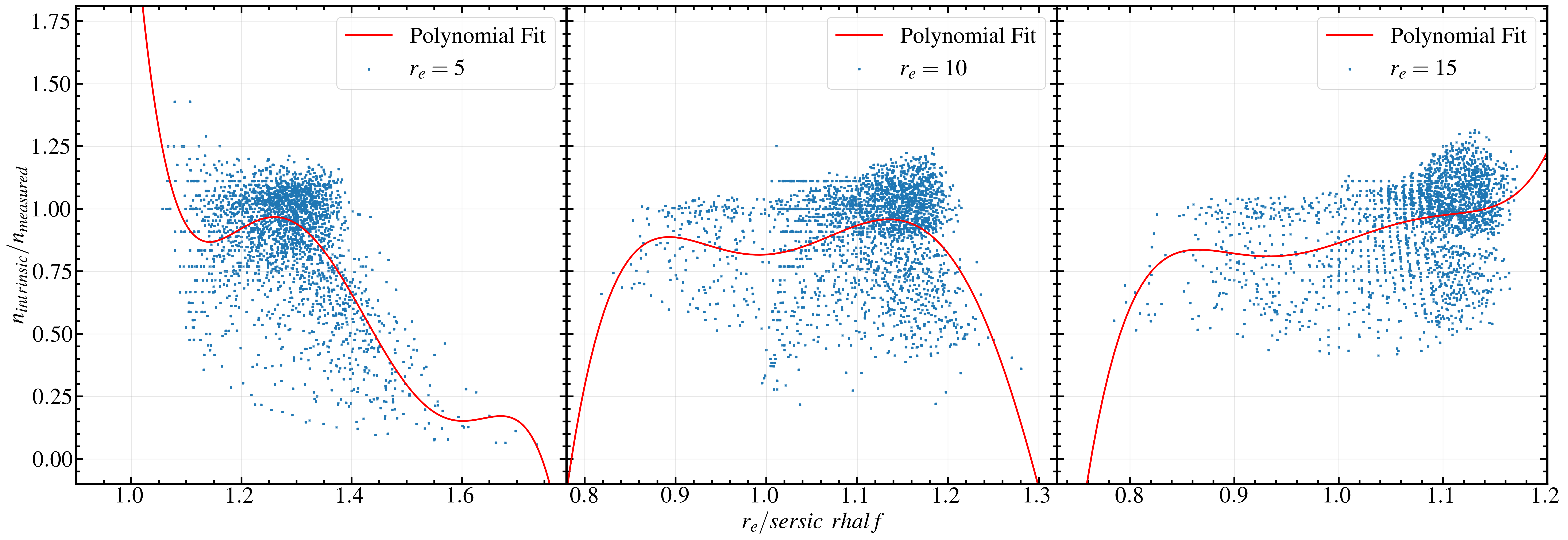}
\caption{The fraction of the intrinsic S\'ersic index $n_{\rm intrinsic}$ to the measured value $n_{\rm measure}$ as a function of the effective radius $r_{\rm e}$ to the effective radius of the 2D S\'ersic profile $sersic\_rhalf$. 
From left to right, the preset effective radii are 5, 10, 15 pixels, respectively. \label{fig:ncorr}}
\end{center}
\end{figure}

In this appendix we describe how we correct the measured S\'ersic index $n$ via the modeling results in \S\ref{sec:correction}.
In Fig.~\ref{fig:ncorr}, we plot the fraction of the intrinsic S\'ersic index $n_{\rm intrinsic}$ to the measured value $n_{\rm measure}$ against the effective radius $r_{\rm e}$ to the effective radius of the 2D S\'ersic profile.
We then performed polynomial fitting to the 5th degree to find how $n_{\rm intrinsic}/n_{\rm measure}$ and $r_{\rm e}/sersic\_rhalf$ are correlated for different radii.

For real objects with $r_{\rm e}<8$, the polynomial fitting for models with $r_{\rm e}=5$ is applied:
\begin{equation}
    \frac{n_{\rm intrinsic}}{n_{\rm measure}}=-384.7x^5 + 2729x^4 - 7679x^3 + 10710x^2 - 7408x + 2033,
\end{equation}
where $x=r_{\rm e}/sersic\_rhalf$ lies within $1.05\leq x\leq 1.70$;

for real objects with $8\leq r_{\rm e}<12$, the polynomial fitting for models with $r_{\rm e}=10$ is applied:
\begin{equation}
        \frac{n_{\rm intrinsic}}{n_{\rm measure}}=893.1x^5 - 4989x^4 + 11020x^3 - 12030x^2 - 6507x - 1394,
\end{equation}
where $x=r_{\rm e}/sersic\_rhalf$ lies within $0.82\leq x\leq 1.27$;

and for real objects with $12\leq r_{\rm e}$, the polynomial fitting for models with $r_{\rm e}=10$ is applied:
\begin{equation}
    \begin{split}
        \frac{n_{\rm intrinsic}}{n_{\rm measure}}=2225x^5 - 11200x^4 +22450x^3 - 22420x^2+11150x -2208,
    \end{split}
\end{equation}
where $x=r_{\rm e}/sersic\_rhalf$ lies within $0.78\leq x\leq 1.18$.

As described in \S\ref{sec:correction}, the $r_{\rm e}$ and $sersic\_rhalf$ calculated for each real object and if the fraction of these two parameters are out of the range of $x$ for each equation, there will be no correction available.
For valid objects, the S\'ersic index can be corrected via simply multiplying the polynomial as a function of $r_{\rm e}/sersic\_rhalf$ by the measured S\'ersic index.

\end{document}